\documentclass[final]{siamonline0316}

\usepackage{amsmath}
\usepackage{lipsum}
\usepackage{amsfonts}
\usepackage{graphicx}
\usepackage{epstopdf}
\usepackage{algorithmic}
\usepackage[caption=false]{subfig}
\ifpdf
  \DeclareGraphicsExtensions{.eps,.pdf,.png,.jpg}
    \graphicspath{{./}}
\else
  \DeclareGraphicsExtensions{.eps,.png,.jpg,.pdf}
  \graphicspath{./}
\fi
\newcommand{\TheTitle}{A forward-adjoint operator pair
based on the elastic wave equation for use in transcranial photoacoustic
computed tomography}

\newcommand{\Theshorttitle}{Discrete elastic wave-equation-based adjoint operator}
\newcommand{\TheAuthors}{K. Mitsuhashi, J. Poudel,
 T.P. Matthews, A. Garcia-Uribe, L.V. Wang and M.A. Anastasio}

\headers{\Theshorttitle}{\TheAuthors}

\title{{\TheTitle}}

\author{
  Kenji Mitsuhashi\and Joemini Poudel
  \and Thomas P. Matthews\and Alejandro Garcia-Uribe\and Lihong V. Wang \thanks{Department of Biomedical Engineering, Washington University in St.Louis, St.Louis, MO USA 
  	(\email{lhwang@wustl.edu}).}
  \and
  Mark A. Anastasio \thanks{Department of Biomedical Engineering, Washington University in St.Louis, St.Louis, MO USA
    (\email{anastasio@wustl.edu}).}
}

\usepackage{amsopn}
\DeclareMathOperator{\diag}{diag}

\newcommand{\V}{\ensuremath{\mathbf{v}}}

\newcommand{\WOp}{\ensuremath{\mathbf{W}}}

\makeatletter
\newcommand{\removelatexerror}{\let\@latex@error\@gobble}
\makeatother
\begin{document}

\maketitle

\begin{abstract}

  Photoacoustic computed tomography (PACT) is an emerging imaging modality that exploits optical contrast and ultrasonic detection principles to form images of the photoacoustically induced initial pressure distribution within tissue. The PACT reconstruction problem corresponds to an inverse source problem
in which the initial pressure distribution
 is recovered from measurements of the radiated wavefield.
 A major challenge in transcranial PACT brain imaging is compensation for aberrations in the measured data due to the presence of the skull. Ultrasonic waves undergo
 absorption, scattering and longitudinal-to-shear wave mode conversion as they propagate through the skull. To properly account for these effects, a wave-equation-based inversion method should be employed that can model the heterogeneous elastic properties of the skull.
 In this work, a forward model based on a finite-difference time-domain discretization of the three-dimensional elastic wave equation is
 established and a procedure for computing the corresponding adjoint of the forward operator is presented. Massively parallel implementations of these operators employing multiple graphics processing units (GPUs) are also developed.
The developed numerical framework is validated and investigated in computer-simulation and experimental
phantom studies whose designs are motivated by transcranial PACT applications.
\end{abstract}



\section{Introduction}
Photoacoustic computed tomography (PACT) is a noninvasive imaging modality that exploits the optical absorption contrast of tissue with the high spatial resolution of ultrasound imaging techniques~\cite{Kruger95,Kruger99,Xu11edema}. In PACT, the target is illuminated with a short optical pulse that results
 in the generation of acoustic pressure signals via the thermoacoustic effect~\cite{Xu06,Oraevsky03}. The propagated acoustic pressure signals are then detected by use of a collection of wideband ultrasonic transducers that are located outside the support of the object. Typically, the measured pressure signals are employed to estimate the induced initial pressure distribution or, equivalently, if the Gr\"{u}neisen parameter is known, the absorbed optical energy distribution.

Transcranial brain imaging represents an important application that may benefit
 significantly by the development of PACT methods.
 Existing human brain imaging modalities
 include X-ray computed tomography (CT),
 magnetic resonance imaging (MRI), positron emission tomography (PET), and
ultrasonography.
 However, all these modalities suffer from significant shortcomings.
X-ray CT, PET, and MRI are expensive and employ
 bulky and generally non-portable imaging equipment.
Moreover, X-ray CT and PET employ ionizing radiation
 and are therefore not suitable for longitudinal studies and MRI-based methods
are generally slow.
 Ultrasonography is
an established portable pediatric brain imaging modality that can operate
in near real-time, but its image quality degrades
severely when employed after the closure of the fontanels.
On the other hand, PACT can be implemented in near real-time, does not employ
ionizing radiation, is much less costly than either MRI, PET or X-ray CT,
and can provide both anatomical and functional information.

In vivo transcranial PACT studies have revealed structure and hemodynamic responses in small animals~\cite{Wang03,Li10,Xu11edema}. In these small animal studies, PACT was used to visualize brain structure, brain lesions and neurofunctional activities such as cerebral hemodynamic responses to hyperoxia and hypoxia, and cerebral cortical responses to various forms of stimulation~\cite{Wang03,Li10}. In addition to providing functional and structural information about the brain, transcranial PACT has also been utlized to study the formation of cerebral edema and its expansion and recovery~\cite{Xu11edema}. Hence, transcranial PACT is a neuroimaging modality that holds promise for applications in neurophysiology, neuropathology and neurotherapy.
Because the skulls in these small animal studies  were relatively thin ($\sim$ 1 $\text{mm}$), they did not significantly aberrate
 the photoacoustic wavefields. As such, conventional backprojection (BP)
 methods that ignored the presence of the skull
and assumed a homogeneous lossless fluid medium were employed for image reconstruction with good success. However, PA signals can be significantly aberrated by thicker skulls present in adolescent and adult primates. To render PACT an effective imaging modality for use with transcranial imaging in large primates, it is necessary to develop image reconstruction methodologies that can accurately compensate for skull-induced aberrations of the recorded PA signals. 

Towards this goal, ex vivo studies involving primate heads have also been conducted~\cite{Xu06monkey,Jin08,Xu11edema,Yang08,Nie11,ChaoSkull}. In such applications, the effects
 of the skull on the 
recorded photoacoustic wavefield were no longer negligible.
To address this,  a subject-specific imaging model that
approximately describes the interaction of the photoacoustic wavefield with the skull
was developed \cite{ChaoSkull}.
This imaging model was established by use of adjunct imaging data, such as X-ray CT, that
specified the skull morphology and composition \cite{Fink,jones2013transcranial,aubry2003experimental}.
It was demonstrated that image reconstruction
based on the subject-specific imaging model  yielded images that contained
significantly reduced artifact levels compared to those reconstructed by use of a conventional BP method \cite{ChaoSkull}. However, a limitation of that work
was that it assumed a fluid medium and therefore assumed a simplified wave propagation
model in which longitudinal-to-shear-wave mode conversion within the skull 
\cite{BobShearJOSA,white,Fry78}, which is an elastic solid, was neglected.  The deleterious effects of making such an approximation in 
transcranial PACT have been studied previously \cite{BobShear}.



To circumvent limitations of previous approaches, in this work a numerical framework
for image reconstruction in transcranial PACT based on an elastic wave equation that describes
an linear isotropic, lossy and heterogeneous medium is developed and investigated.
Similar to the work by Huang \emph{et al.} \cite{ChaoSkull}, estimates of the acoustic parameters of the skull
are assumed to be obtainable from adjunct image data.
In transcranial ultrasound therapy applications, the skull's acoustic parameters are routinely
estimated in this way \cite{Fink,jones2013transcranial,aubry2003experimental,marquet2009non}.
However, unlike previous methods, the skull is treated as an elastic solid
and therefore longitudinal-to-shear-wave mode conversion within the skull is modeled.

 The primary contributions of this work are the establishment
 of a discrete forward operator (i.e., imaging model) for transcranial PACT that is based
 on the three-dimensional (3D) elastic wave equation and a procedure to implement an associated matched adjoint operator.
 Specifically, the finite-difference time-domain method (FDTD) is adopted for implementing the forward operator.
 Both the  forward and adjoint operators are implemented using multiple graphics processing units (GPUs).
In certain cases, the adjoint operator may serve as a useful image reconstruction operator.  More generally, however, the ability to
compute the adjoint operator will permit application of gradient-based iterative reconstruction algorithms that seek to minimize a specified objective function.
The developed numerical framework is validated and investigated in computer-simulation and experimental
phantom studies.

The paper is organized as follows. In Section~\ref{sec:background}, the salient
 imaging physics and image reconstruction principles are reviewed
briefly. The explicit formulation of the forward operator is described in Section~\ref{sec:discrete},
 and the implementation of the corresponding discrete adjoint operator
 is described in Section~\ref{sec:Description}.  The forward operator is validated
in Section~\ref{sec:validate}, in which simulated and analytically-produced measurement
data are compared.
Finally, the forward and adjoint operators are applied in image reconstruction studies
involving computer-simulated and experimental data in  Sections~\ref{sec:three} and~\ref{sec:Results},
respectively.

\section{Background}
\label{sec:background}

The principles of photoacoustic wavefield generation and propagation in an elastic medium are
described below in their continuous and discrete forms.
 The discrete description is based on the FDTD method~\cite{Boore72,Moczo07,Moczorobertson07,Virieux86}.
The FDTD method is described by use of a matrix notation, which subsequently will facilitate the computation of the matched adjoint operator.

\subsection{Photoacoustic wavefield propagation: Continuous formulation}
Let the \\ \mbox{photoacoustically-induced} stress tensor at location $\mathbf{r} \in \mathbb{R}^3$ and time $t \geq 0$ be defined as
\begin{align} 
\boldsymbol{\sigma}(\mathbf{r},t)\equiv
\begin{bmatrix}
	&\boldsymbol{\sigma}^{11}(\mathbf{r},t) &\boldsymbol{\sigma}^{12}(\mathbf{r},t) &\boldsymbol{\sigma}^{13}(\mathbf{r},t)\hspace{10pt}
	\\ 
	&\boldsymbol{\sigma}^{21}(\mathbf{r},t) &\boldsymbol{\sigma}^{22}(\mathbf{r},t) &\boldsymbol{\sigma}^{23}(\mathbf{r},t)\hspace{10pt}
	\\
    &\boldsymbol{\sigma}^{31}(\mathbf{r},t) &\boldsymbol{\sigma}^{32}(\mathbf{r},t) &\boldsymbol{\sigma}^{33}(\mathbf{r},t)\hspace{10pt} 
\end{bmatrix},
\end{align} 
where $\boldsymbol{\sigma}^{ij}(\mathbf{r},t)$ represents the stress in the $i^{\text{th}}$ direction acting on a plane perpendicular to the $j^{\text{th}}$ direction. Additionally, let $p_0(\mathbf{r})$ denote the photoacoustically-induced initial pressure distribution  within the object, and  $\dot{\mathbf{u}}(\mathbf{r}, t)$ $\equiv$ ($\dot{u}^1(\mathbf{r},t), \dot{u}^2(\mathbf{r},t), \dot{u}^3(\mathbf{r},t)$) represent the vector-valued acoustic particle velocity. Let $\rho(\mathbf{r})$ denote medium's density distribution and $\lambda(\mathbf{r})$, $\mu(\mathbf{r})$  represent the Lam\'e parameters that describe the full elastic tensor of the linear isotropic media.
All functions in this work are assumed to be bounded and compactly supported.

 The compressional and shear wave propagation speeds are given by, 
\begin{subequations} 
	\begin{align}
		c_l(\mathbf{r}) = \sqrt{\frac{\lambda(\mathbf{r}) + 2 \mu(\mathbf{r})}{\rho(\mathbf{r})}} \text{ and }
		c_s(\mathbf{r}) = \sqrt{\frac{\mu(\mathbf{r})}{\rho\left(\mathbf{r}\right)	}},
	\end{align}
\end{subequations}
respectively.
In transcranial PACT imaging applications, the acoustic absorption is not negligible. Here,  acoustic absorption within the skull is described by a diffusive absorption
 model~\cite{pinton2011}. The diffusive absorption model ignores the
fact that the wavefield absorption is dependent on temporal frequency.
 This model, however, is reasonable for cases where the bandwidth of the photoacoustic signals are limited. Moreover, the model also assumes that the shear absorption to compressional absorption ratio is given by the compressional velocity to shear velocity ratio. This is approximately true in bone, because the slower shear waves are, in fact, more attenuated than the faster compressive waves~\cite{pinton2011}. 

In a 3D heterogeneous linear isotropic elastic medium with an acoustic absorption coefficient $\alpha(\mathbf{r})$, the propagation of $\dot{\mathbf{u}}(\mathbf{r},t)$ and $\boldsymbol{\sigma}(\mathbf{r},t)$ can be modeled by the following two coupled equations~\cite{Boore72,Virieux86,Madariaga98,AltermanandKaral1968}:
\begin{subequations}
\begin{align}
\partial_t \dot{\mathbf{u}}\left(\mathbf{r},t\right) + \alpha\left(\mathbf{r}\right) \dot{\mathbf{u}}\left(\mathbf{r},t\right)= \frac{1}{\rho\left(\mathbf{r}\right)}\Big(\nabla \cdot \boldsymbol{\sigma}\left(\mathbf{r},t\right)\Big)
\label{eq:subeq1}
\end{align}
and
\begin{align}
  \partial_t \boldsymbol{\sigma}\left(\mathbf{r},t\right) = \lambda(\mathbf{r}) \textbf{tr}(\nabla\dot{\mathbf{u}}\left(\mathbf{r},t\right)) \mathbf{I} +  \mu(\mathbf{r})(\nabla \dot{\mathbf{u}} \left(\mathbf{r},t\right) + \nabla \dot{\mathbf{u}}\left(\mathbf{r},t\right) ^ {T}),
  \label{eq:subeq2}
\end{align}
\text{subject to the initial conditions}
\begin{align}\label{eq:Ini}
	\boldsymbol{\sigma}_0(\mathbf{r}) \equiv \boldsymbol{\sigma}(\mathbf{r},t)|_{t = 0} = -\frac{1}{3} p_0(\mathbf{r})\mathbf{I}, \ \ \ \ \ \ \  \dot{\mathbf{u}}\left(\mathbf{r},t\right)|_{t=0} = 0.
\end{align}
\label{eq:Elastic}
\end{subequations} 
Here, $\textbf{tr}$ $(\cdot)$ is the operator that calculates the trace of a matrix and $\mathbf{I}\in \mathbb{R}^{3 \times 3}$ is the identity matrix.
 In Eqn.~\cref{eq:Ini}, it has been assumed that the object function $p_0(\mathbf{r})$ is compactly supported in a fluid medium where the shear modulus $\mu(\mathbf{r}) = 0 $.  In transcranial PACT, this corresponds to the situation
where the initial photoacoustic wavefield is produced within
the soft tissue enclosed by the skull. Note, the initial conditions defined in Eqn.~\cref{eq:Elastic} would have to be modified if the photoacoustically-induced stress is generated within the elastic media. 

\subsection{Photoacoustic wavefield propagation: Discrete formulation}
\label{sec:FDTD}

The FDTD method is employed to propagate the photoacoustic wavefield forward in space and time  by computing numerical solutions to the coupled equations
 given in Eqn.~\cref{eq:Elastic}~\cite{Moczo07}.
In the FDTD method, at a given temporal step, each grid point is updated based on the local information around that same point. As the
 FDTD method utilizes local information, it lends itself to distributed programming across multiple devices. Since the latency in communication between devices is a limiting factor on computational efficiency,
 spectral and pseudo-spectral methods ~\cite{Carcione94,Tessmer94,Treeby,Treeby2014}  that use global information at each temporal update are not as efficient for distributed programming. In addition, because of the simplicity of the FDTD method to model elastic wave propagation, it still remains very widely employed in seismology~\cite{Moczo07}.

\begin{figure}[htbp]
	\centering  \includegraphics[height=2.0in]{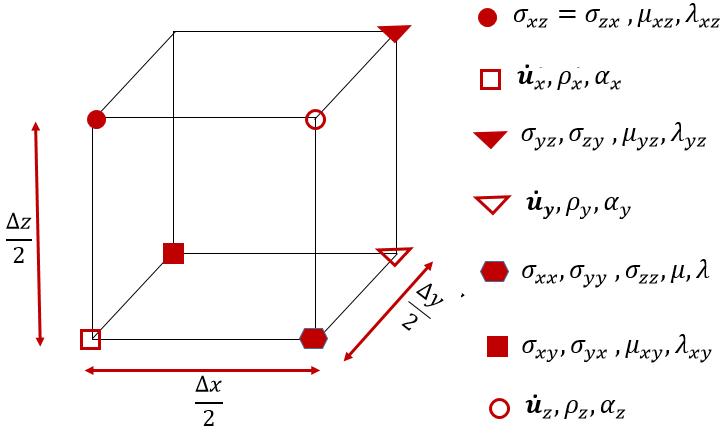}  \centering
	\caption{A staggered-grid FD cell with positions of the wavefield variables~\cite{Moczo07}.}
	\label{fig:staggered}
\end{figure}

The salient features of the FDTD method that will underlie the discrete PACT imaging model are described below.
 In the initial applications of the FDTD method
 to the elastic wave equation, all functions (e.g., stress tensor and particle velocity) were sampled at the same grid positions~\cite{AltermanandKaral1968,Boore72}. However, such conventional grid schemes were found to possess
 limitations. Problems with grid dispersion and instabilities in media
 possessing high Poisson's ratios led Virieux~\cite{Virieux86,Virieux84} to introduce the staggered-grid velocity-stress finite difference schemes for modeling elastic wave propagation. In our study we will employ the introduced the $4^{th}$-order staggered-grid FD scheme to compute the numerical spatial derivatives. The $4^{th}$-order staggered-grid scheme  in 3D has been shown to reduce the computer memory requirements by at least
 eight times compared to $2^{nd}$-order schemes without any loss in accuracy~\cite{Moczo07}. A staggered-grid FD cell with positions where the particle-velocity components, stress-tensor components, density, elastic material parameters are sampled is illustrated in~\cref{fig:staggered}~\cite{Moczo07}.

Note that for the FD scheme, the material properties, stress, and particle velocity functions, are sampled at different points of the staggered FD cell depicted in~\cref{fig:staggered}. Let the set of position vectors  $\{\mathbf{r}^i_1, \mathbf{r}^i_2,..., \mathbf{r}^i_N \in \mathbb{R}^3\}$ specify the locations where $u^i,\text{ for } i=1,2,3$, is sampled and $\{\mathbf{r}^{jk}_1, \mathbf{r}^{jk}_2,..., \mathbf{r}^{jk}_N \in \mathbb{R}^3\}$ specify the locations where $\sigma^{jk}(\text{for } j,k = 1,2,3)$ is sampled. Here, $N = N_1N_2N_3$ specifies the number of vertices of a 3D Cartesian grid, where $N_i$ denotes the number of vertices along the $i^{th}$ direction.
Additionally, let $m\Delta t$, $m \in \mathbb{Z}^{*},\ \Delta t \in \mathbb{R}^{+}$, denote discretized values of the temporal coordinate $t$,
 where $\mathbb{Z}^{*}$ and $\mathbb{R}^{+}$ denote the sets of non-negative integers and positive real numbers, respectively.  Lexicographically ordered vector representations of the components of the sampled particle velocity and stress tensor will be denoted as 
  \begin{subequations}\label{eq:sigu}
  \begin{align}
  &\dot{\mathbf{u}}_m^i = \left[\dot{\mathbf{u}}^i(\mathbf{r}_1^i,m\Delta t), ..., \dot{\mathbf{u}}^i(\mathbf{r}_N^i,m\Delta t)\right]^T \label{eq:vectoru} \\
  &\text{and} \nonumber \\
  &\boldsymbol{\sigma}_m^{jk} = \left[\sigma^{jk}(\mathbf{r}_1^{jk},m\Delta t), ..., \sigma^{jk}(\mathbf{r}_N^{jk},m\Delta t)\right]^T .\label{eq:sstress}
  \end{align}
  \end{subequations} 
  Let $\mathbf{Q}_i\text{, }\mathbf{A}_{i}\text{, }\boldsymbol{\Lambda}_{jk}\text{ and } \mathbf{M}_{jk} \in \mathbb{R}^{N \times N}$ be matrices describing the elastic properties of the medium defined as
  \begin{subequations}
  \begin{align}
  &\mathbf{Q}_i \equiv \diag\left[\rho(\mathbf{r}_1^i),...,\rho(\mathbf{r}_N^i)\right], \\
  &\mathbf{A}_{i} \equiv \diag\left[\alpha(\mathbf{r}_1^{i}),...,\alpha(\mathbf{r}_N^{i})\right], \\
  &\boldsymbol{\Lambda}_{jk} \equiv \diag\left[\lambda(\mathbf{r}_1^{jk}),...,\lambda(\mathbf{r}_N^{jk})\right], \\
  &\text{ and } \nonumber \\
  &\boldsymbol{\mathcal{M}}_{jk} \equiv \diag\left[\mu(\mathbf{r}_1^{jk}),...,\mu(\mathbf{r}_N^{jk})\right],
  \end{align}
  \end{subequations}
  where $\diag(a_1,...,a_N)$ denotes a diagonal matrix whose diagonal entries starting in the upper left corner are $a_1,...,a_N$.
  	Let $\overline{\dot{\mathbf{u}}}_{m-\frac{1}{2}} \in \mathbb{R}^{3N \times 1}$ and $\overline{\boldsymbol{\sigma}}_m \in \mathbb{R}^{6N \times 1}$ be the concatenation of the unique components of the particle velocity and stress tensor, defined as
  	\begin{subequations}
  		\begin{align}
  		\overline{\dot{\mathbf{u}}}_{m - \frac{1}{2}} &\equiv \left[\dot{\mathbf{u}}_{m - \frac{1}{2}}^1, \dot{\mathbf{u}}_{m - \frac{1}{2}}^2, \dot{\mathbf{u}}_{m - \frac{1}{2}}^3\right]^T \\
 &\text{ and } \nonumber \\
  		\overline{\boldsymbol{\sigma}}_m &\equiv \left[\boldsymbol{\sigma}_m^{11},  \boldsymbol{\sigma}_m^{22}, \boldsymbol{\sigma}_m^{33}, \boldsymbol{\sigma}_m^{23}, \boldsymbol{\sigma}_m^{13}, \boldsymbol{\sigma}_m^{12}\right]^T.
  		\end{align}
  	\end{subequations}
  		Note that because of the symmetry of the stress tensor (e.g., $\boldsymbol{\sigma}^{ij} = \boldsymbol{\sigma}^{ji}$), it is not necessary to calculate all nine components of the second-order tensor. Here, we have chosen the ordering of $\overline{\boldsymbol{\sigma}}_m$ to follow
 Voigt notation~\cite{Helnwein01}.
  		
  		To simplify the subsequent presentation, the following operators are defined:
  		\begin{subequations}\label{eq:Operatordef}
  			\begin{align}
  			\mathbf{J}_i \overline{\dot{\mathbf{u}}}_{m - \frac{1}{2}} &\equiv \left(\mathbf{I}_{N \times N} - \Delta t \mathbf{A}_i\right) \dot{\mathbf{u}}_{m - \frac{1}{2}}^i \\
  			\boldsymbol{\Phi}_{i} \overline{\boldsymbol{\sigma}}_m &\equiv \Delta t \mathbf{Q}_i^{-1} \sum_{j=1}^{3} \partial_j \boldsymbol{\sigma}_m^{ij} \\
  			\boldsymbol{\Psi}_{ij} \overline{\dot{\mathbf{u}}}_{m - \frac{1}{2}} &\equiv \Delta t \left[\delta_{ij} \boldsymbol{\Lambda}_{ij} \sum_{k=1}^{3}\partial_k \dot{\mathbf{u}}_{m - \frac{1}{2}}^k + \boldsymbol{\mathcal{M}}_{ij} \left(\partial_i \dot{\mathbf{u}}_{m - \frac{1}{2}}^j + \partial_j \dot{\mathbf{u}}_{m - \frac{1}{2}}^i\right)\right] ,
  			\end{align}
  		\end{subequations}
  		where $\delta_{ij}$ is the Kronecker delta, $\mathbf{J}_i, \boldsymbol{\Psi}_{ij} \in \mathbb{R}^{N \times 3N}$, $\boldsymbol{\Phi}_{i} \in \mathbb{R}^{N \times 6N}$, and $\partial_i$ denotes the partial derivative with respect to the $i^{th}$ spatial coordinate. These operators will allow us to compactly express the discrete form of Eqn.~\cref{eq:Elastic}. In addition, the spatial derivatives defined in Eqn.~\cref{eq:Operatordef} are calculated using a $4^{th}$-order finite-difference scheme. 
  		The $4^{\text{th}}$-order finite-difference approximation to the first derivative along any arbitrary direction x in a staggered grid setup is given by~\cite{Moczo07}
  		\begin{multline}\label{eq:derivative}
  			\frac{d \phi}{dx}(x_0) = \frac{1}{\Delta x}\Bigg[ -\frac{1}{24}\Big( \phi(x_0 + \frac{3}{2}\Delta x) - \phi(x_0 - \frac{3}{2}\Delta x)\Big) + \\
  			\frac{9}{8} \Big( \phi(x_0 + \frac{1}{2}\Delta x) - \phi(x_0 - \frac{1}{2}\Delta x) \Big)\Bigg],
  		\end{multline}
  		where $\phi(x)$ is an arbitrary field variable (e.g., stress tensor or particle velocity for elastic wave equation). 
Additionally, define the operators 
  		\begin{subequations}
  			\begin{align}
  			\mathbf{J} &\equiv \left[\mathbf{J}_1, \mathbf{J}_2, \mathbf{J}_3\right]^T, \\
  			\boldsymbol{\Phi} &\equiv \left[\boldsymbol{\Phi}_1, \boldsymbol{\Phi}_2, \boldsymbol{\Phi}_3\right]^T,\text{ and }\\
  			\boldsymbol{\Psi} &\equiv \left[\boldsymbol{\Psi}_{11}, \boldsymbol{\Psi}_{22}, \boldsymbol{\Psi}_{33}, \boldsymbol{\Psi}_{23}, \boldsymbol{\Psi}_{13}, \boldsymbol{\Psi}_{12}\right]^T,
  			\end{align}
  		\end{subequations}
  		where $\mathbf{J} \in \mathbf{R}^{3N \times 3N}$, and $\boldsymbol{\Phi} \in \mathbb{R}^{3N \times 6N}$, $\boldsymbol{\Psi} \in \mathbb{R}^{6N \times 3N}$.
 In terms of these quantities, the discretized forms of Eqns.~\cref{eq:subeq1} and \cref{eq:subeq2}
can be expressed as
  			\begin{subequations}\label{eq:it}
  				\begin{align}
  				\overline{\dot{\mathbf{u}}}_{m+\frac{1}{2}} &= \mathbf{J} \overline{\dot{\mathbf{u}}}_{m - \frac{1}{2}} + \boldsymbol{\Phi} \overline{\boldsymbol{\sigma}}_m \\
  				\overline{\boldsymbol{\sigma}}_{m+1} &= \overline{\boldsymbol{\sigma}}_{m} + \boldsymbol{\Psi} \overline{\dot{\mathbf{u}}}_{m+\frac{1}{2}}.
  				\end{align}
  			\end{subequations}

\subsection{Specification of the image reconstruction problem}
The PACT image reconstruction problem addressed in this work
 is to obtain an estimate of the photoacoustically induced initial pressure distribution $p_0\left(\mathbf{r}\right)$ from pressure measurements recorded by a collection of ultrasonic transducers surrounding the object. Let $\mathbf{\hat{p}}_m\ \equiv \ (p(\mathbf{r}_0^d,m \Delta t),\cdots,p(\mathbf{r}_{L - 1}^d,m \Delta t))^T$ denote the measured pressure wavefield at time $t = m\Delta t\text{, for } m = 0,\cdots,M-1$, where $M$ is the total number of time steps, and let $\mathbf{r}_l^d \in \mathbb{R}^3 \text{, for }l=0,\cdots,L-1$ denote the positions of the $L$ ultrasonic transducers that reside outside the support of the object $p_0(\mathbf{r})$. Here, for simplicity, we neglect the acousto-electrical impulse response of the ultrasonic transducers and assume each transducer is point-like. However,
 we can incorporate the spatial and electrical impulse responses of the transducers in the developed discrete imaging model~\cite{Huang13}. In addition, the acoustic parameters of the medium are assumed to be known.

A general form of the discrete PACT imaging model can be expressed as 
\begin{align}\label{eq:Imageeqn}
\mathbf{\hat{p}} = \mathbb{H}\mathbf{p}_0,
\end{align}
where the $LM \times 1$ vector 
\begin{align}
\mathbf{\hat{p}} \equiv \begin{bmatrix}
&\mathbf{\hat{p}}_0&\\
&\mathbf{\hat{p}}_1&\\
&\vdots&\\
&\mathbf{\hat{p}}_{M-1}&
\end{bmatrix}
\end{align}
represents the measured pressure data corresponding to all transducer locations and temporal samples. Additionally, $\mathbf{p}_0 \in \mathbb{R}^{N\times 1}$ is the discrete representation of the sought after initial pressure distribution within the object that is given by
\begin{align}\label{eq:p0}
	\mathbf{p}_0 = -\sum_{i} \boldsymbol{\sigma}_0^{ii},
\end{align} 
where $\boldsymbol{\sigma}_0^{jk}$ is the photoacoustically induced initial stress distribution (Eqn.\cref{eq:sstress} with $m = 0$).
The $LM \times N$ matrix $\mathbb{H}$ represents the discrete imaging operator (that specifies the forward model), also referred to as the system matrix. The construction of the system matrix $\mathbb{H}$ is based on the initial value problem defined in Eqn.~\cref{eq:Elastic}, and is described in great detail in Section~\ref{sec:discrete}.

The image reconstruction task in a discrete setting is to determine an estimate of $\mathbf{p}_0$ from knowledge of the measured data $\mathbf{\hat{p}}$.
Multiple classes of iterative image reconstruction algorithms  require
the actions of the operators
 $\mathbb{H}$ and its adjoint $\mathbb{H}^{\dagger}$ to be 
computed repeatedly \cite{boyd2004convex,combettes2011proximal,beck2014introduction}.
Moreover, in some cases, the adjoint operator may serve as a useful heuristic reconstruction
operator.
 Methods for implementing these operators are described below. 

\section{Explicit formulation of discrete imaging model}
\label{sec:discrete}
The FDTD method for numerically solving the photoacoustic wave equation in elastic linear isotropic media described in Section~\ref{sec:FDTD} will be employed to implement the action of the system matrix $\mathbb{H}$. In this section, we provide an explicit representation of $\mathbb{H}$ that will subsequently be employed to determine $\mathbb{H}^{\dagger}$. 
Equation~\cref{eq:it} can be described by a single matrix equation to determine the updated wavefield variables after a time step $\Delta t$ as
\begin{equation}
\V_{m+1}=\WOp\V_m,
\end{equation} 
where  
\begin{align}
		\mathbf{v}_m = \begin{bmatrix}
		\overline{\dot{\mathbf{u}}}_{m - \frac{1}{2}} \\
		\overline{\boldsymbol{\sigma}}_m
		\end{bmatrix},
  \end{align} 
and $\mathbf{W} \in \mathbb{R}^{9N\times 9N}$ is the propagator matrix defined as
	\begin{align}\label{eq:W}
	\mathbf{W} \equiv \begin{bmatrix}
	\mathbf{J} & \boldsymbol{\Phi} \\
	\boldsymbol{\Psi}\mathbf{J} & \mathbf{I}_{6N \times 6N} + \boldsymbol{\Psi} \boldsymbol{\Phi} 
	\end{bmatrix}.
	\end{align}
The wavefield quantities can be propagated forward in time from $t = 0$ to $t = (M-1)\Delta t$ as 
\begin{align}\label{eq:Propagate}
\begin{bmatrix}
&\mathbf{v}_0&\\
&\mathbf{v}_1&\\
&\vdots&\\
&\mathbf{v}_{M-1}&\\
\end{bmatrix}
= \mathbf{T}_{M-1}\cdots\mathbf{T}_{1}
\begin{bmatrix}
&\mathbf{v}_0&\\
&\mathbf{0}_{9N \times 1}&\\
&\vdots&\\
&\mathbf{0}_{9N \times 1}&\\
\end{bmatrix},
\end{align}
where the $9NM \times 9NM$ matrices $\mathbf{T}_m (m= 1,...,M-1)$ are defined in terms of $\mathbf{W}$ as 
\begin{align}
\mathbf{T}_m \equiv 
\begin{bmatrix}
\mathbf{I}_{9N \times 9N}&\cdots&\mathbf{0}_{9N \times 9N}&\ \ \\
\vdots&\ddots&\vdots&\mathbf{0}_{(m+1)\cdot9N \times (M-m)\cdot9N}\\
\mathbf{0}_{9N \times 9N}&\cdots&\mathbf{I}_{9N \times 9N}&\ \ \\
\mathbf{0}_{9N \times 9N}&\cdots&\mathbf{W}&\ \ \\
\mathbf{0}_{(M-m-1)\cdot9N \times m\cdot9N}&\ &\ &\mathbf{0}_{(M-m-1)\cdot9N \times (M-m)\cdot9N}
\end{bmatrix}
\end{align}
with $\mathbf{W}$ residing between the $(9N(m-1) + 1)^{th}$ to $9Nm^{th}$ columns and the $(9Nm + 1)^{th}$ to $9N(m+1)^{th}$ rows of $\mathbf{T}_m$.

From the initial conditions in Eqn.~\cref{eq:Ini}, the vector $(\mathbf{v}_0,\mathbf{0}_{9N \times 1},\cdots,\mathbf{0}_{9N \times 1})^T$ can be computed from the initial pressure distribution $\mathbf{p}_0$ as 
\begin{align}\label{eq:initialp0}
\begin{bmatrix}
&\mathbf{v}_0&\\
&\mathbf{0}_{9N \times 1}&\\
&\vdots&\\
&\mathbf{0}_{9N \times 1}&\\
\end{bmatrix} =
\mathbf{T}_0\mathbf{p}_0
,
\end{align}
\text{ where }
\begin{align}
\mathbf{T}_0 &\equiv
\begin{bmatrix}
	\boldsymbol{\tau},	
	\mathbf{0}_{9N\times N},
	\cdots,
	\mathbf{0}_{9N\times N}	
\end{bmatrix}^T \in \mathbb{R}^{9NM\times N} \\
&\text{and}\nonumber\\ 
\boldsymbol{\tau} &\equiv 
\begin{bmatrix}
	\mathbf{0}_{3N\times N},
	-\frac{1}{3}\mathbf{I}_{N\times N},
	-\frac{1}{3}\mathbf{I}_{N\times N},
	-\frac{1}{3}\mathbf{I}_{N\times N},
	\mathbf{0}_{3N\times N}	
\end{bmatrix}^T \in \mathbb{R}^{9N\times N},
\end{align}
with $\mathbf{p}_0$ being specified by Eqn.~\cref{eq:p0}. 

In general, the transducer locations $\mathbf{r}_l^d$ at which the PA data $\mathbf{\hat{p}}$ are recorded will not coincide with the vertices of the 3-D Cartesian grid at which the propagated field quantities are computed. The measured data $\mathbf{\hat{p}}$ can be related to the computed field quantities via an interpolation operation defined as 
\begin{align}\label{eq:interpgrid}
\mathbf{\hat{p}} = \mathbf{M} \begin{bmatrix}
&\mathbf{v}_0&\\
&\mathbf{v}_{1}&\\
&\vdots&\\
&\mathbf{v}_{M -1}&\\
\end{bmatrix}, 
\text{ where } 
\mathbf{M} \equiv \begin{bmatrix}
\boldsymbol{\Theta}&\mathbf{0}_{L \times 9N}&\cdots&\mathbf{0}_{L \times 9N}\\
\mathbf{0}_{L \times 9N}&\boldsymbol{\Theta}&\cdots&\mathbf{0}_{L \times 9N}\\
\vdots&\vdots&\ddots&\vdots\\
\mathbf{0}_{L \times 12N}&\mathbf{0}_{L \times 9N}&\cdots&\boldsymbol{\Theta}
\end{bmatrix} \in \mathbb{R}^{LM \times 9NM}.
\end{align}
Here, $\boldsymbol{\Theta} \equiv \begin{bmatrix}
s_1,\cdots,s_L
\end{bmatrix}^T \in \mathbb{R}^{L\times 9N}$, where $l= 1,\cdots,L$ and 
\begin{align}
s_l = \begin{bmatrix} \mathbf{0}_{1\times 3N}, -\mathbf{R}_l,-\mathbf{R}_l, -\mathbf{R}_l, \mathbf{0}_{1\times 3N} \end{bmatrix}
\end{align}
 is a $1  \times 9N$ row vector. The elements of
the row vector $\mathbf{R}_l \in \mathbb{R}^{1\times N}$
 are assigned values to compute the pressure wavefield at the $l^{th}$ transducer using trilinear interpolation.
 
By use of Eqns.~\cref{eq:Propagate},~\cref{eq:initialp0}, and~\cref{eq:interpgrid},
the PACT imaging model in Eq.\ (\ref{eq:Imageeqn}) can be expressed as
\begin{align}\label{eq:STforH}
\mathbf{\hat{p}} = \mathbf{M}\mathbf{T}_{M-1}\cdots\mathbf{T}_1\mathbf{T}_0\mathbf{p}_0.
\end{align}
and therefore the system matrix is identified as
\begin{align}\label{eq:H}
\mathbb{H} = \mathbf{M}\mathbf{T}_{M-1}\cdots\mathbf{T}_1\mathbf{T}_0.
\end{align}
 The explicit form of $\mathbb{H}^{\dagger}$ is therefore given by
\begin{align}\label{eq:adjH}
\mathbb{H}^{\dagger} = \mathbf{T}_0^{\dagger}\mathbf{T}_1^{\dagger}\cdots \mathbf{T}_{M-1}^{\dagger}\mathbf{M}^{\dagger},
\end{align}
where the superscript $\dagger$ denotes the conjugate transpose of a matrix. 

\section{Implementation of the forward and adjoint operators}
\label{sec:Description}
Since a typical computational grid for simulating the stress tensor field that covers an entire human skull can
 consist of 500 million cells or more, there is a need for computationally
efficient implementations of the forward and adjoint operators.
To address this, a massively parallel implementation of the FDTD method
 based on NVIDIA's CUDA framework for general-purpose GPU
 computation was implemented \cite{Michea10,Micikevicius09}.
In this way, the action of both the operators $\mathbb{H}$ and $\mathbb{H}^{\dagger}$
 could be computed by use of multiple GPUs utilizing the message passing interface (MPI)~\cite{Micikevicius09}.

To prevent acoustic waves from reflecting off the edge of the simulation grid, an anisotropic absorbing boundary condition called a perfectly matched layer (PML) was implemented~\cite{Berenger94,Berenger96,Berenger96a,Berenger97}. For solving the photoacoustic wave equation in elastic, linear isotropic media, a special form of the PML called a convolutional-PML (C-PML) was implemented~\cite{Komatitsch07,Roden00}. To incorporate the C-PML, auxiliary memory variables need to be introduced~\cite{Komatitsch07,Roden00}. Due to the incorporation of the auxiliary memory variables, both $\mathbb{H}$ and $\mathbb{H}^{\dagger}$ need to be modified. The modified $\mathbb{H}$ and $\mathbb{H}^{\dagger}$ operators after the incorporation of the C-PML are described in the Appendix.

The action of the adjoint matrix was implemented according to Eqn.~\cref{eq:adjH}. It can be verified that $\mathbf{p}^{adj} = \mathbb{H}^{\dagger}\mathbf{\hat{p}}$ can be computed as
\begin{subequations}
\begin{align}
\mathbf{v}_{M-1} &= \boldsymbol{\Theta}^{T}\mathbf{\hat{p}}_{M-1}\\
\mathbf{v}_{m - 1} &= \boldsymbol{\Theta}^{T}\mathbf{\hat{p}}_{m -1} + \mathbf{W}^{T}\mathbf{v}_{m} \nonumber \\
&m = M-1,\cdots,1 \label{eq:Adjup}\\
\mathbf{p}^{adj} &= \boldsymbol{\tau}^{T}\mathbf{v}_0.
\end{align}
\end{subequations}
Moreover, the recursive temporal backward update step of Eqn.~\cref{eq:Adjup} can be written in terms of the update of the field variables similar to Eqn.~\cref{eq:it} as 
	\begin{subequations}\label{eq:backwardstep}
		\begin{align}
		\tilde{\dot{\mathbf{u}}}_{m - \frac{1}{2}} &= \overline{\dot{\mathbf{u}}}_{m+\frac{1}{2}} + \boldsymbol{\Psi}^T \overline{\boldsymbol{\sigma}}_{m + 1} \\
		\overline{\boldsymbol{\sigma}}_m &= \overline{\boldsymbol{\sigma}}_{m + 1} + \boldsymbol{\Phi}^T \tilde{\dot{\mathbf{u}}}_{m - \frac{1}{2}} + \boldsymbol{\mathcal{I}}_2 \boldsymbol{\Theta}^T \hat{\mathbf{p}}_{m+1} \\
		\overline{\dot{\mathbf{u}}}_{m - \frac{1}{2}} &= \mathbf{J} \tilde{\dot{\mathbf{u}}}_{m - \frac{1}{2}}
		\end{align}
	\end{subequations}
	where $\boldsymbol{\mathcal{I}}_2 \mathbf{v}_m \equiv \overline{\boldsymbol{\sigma}}_m$.

\section{Validation studies}
\label{sec:validate}
\begin{figure}[!t]
	\centering
	\subfloat[]{\includegraphics[height=2.00in, width = 3.00in]{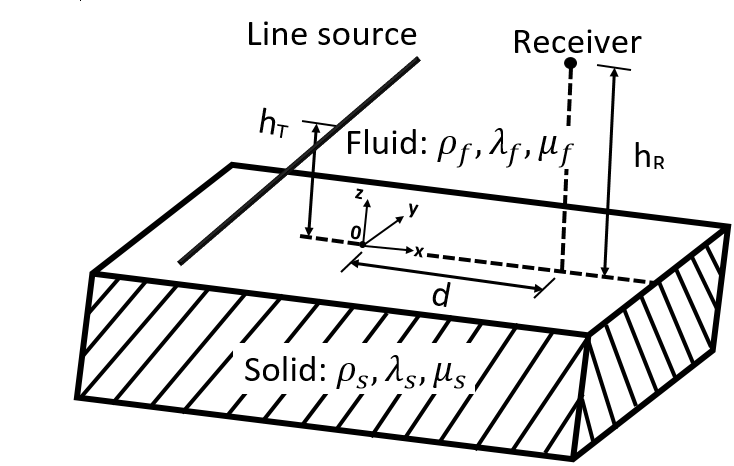}%
		\hspace{0.01in}
		\label{fig:validate}}
	\subfloat[]{\includegraphics[height=2.00in, width = 3.00in]{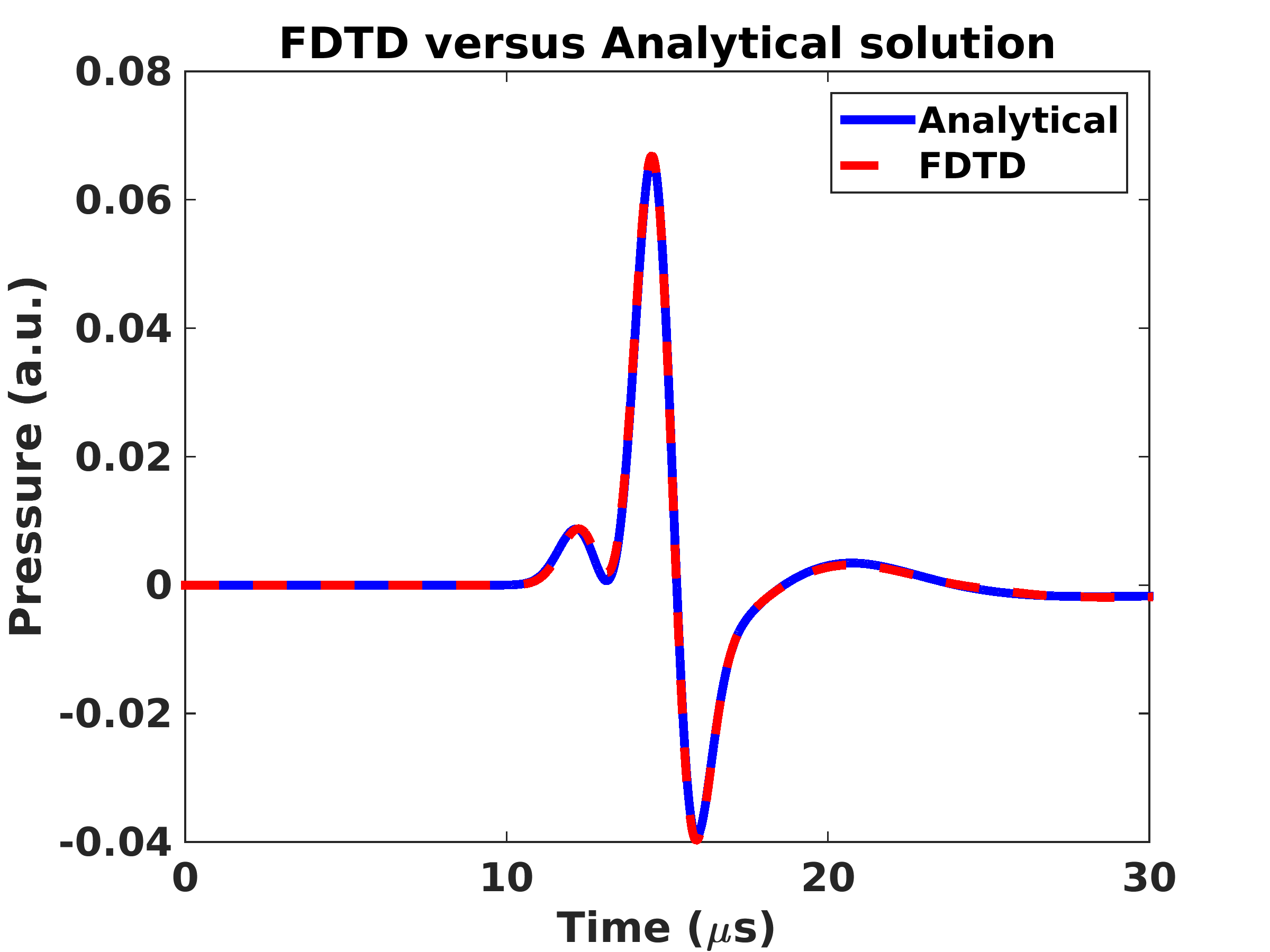}%
		\label{fig:deHoop}}
	\caption{(a) The setup used to conduct the validation study. (b) A plot comparing the pressure profile obtained at the receiver location using the analytical solution and the FDTD simulation.}
\end{figure}

The implementation of the forward operator $\mathbb{H}$,
 as described by Eqn.~\cref{eq:H}, was validated by comparing the results obtained from the FDTD simulation with a known analytical solution. As shown in~\cref{fig:validate}, the analytic solution was computed for a lossless semi-infinite medium, with fluid and linear isotropic solid media divided by a planar boundary. A monopole line source was placed in the fluid medium with the $z$-axis chosen normal to the interface between the solid/fluid media. Furthermore, the $y$-axis was chosen parallel to the line source located at $x = 0$, $z = h_T$. The receiving transducer, in this configuration, was
 located at $x = d$, $z= h_R$. In the setup described, the strength of the line source and the properties of the configuration were both independent of $y$. The analytical solution was computed via the Cagniard-De Hoop method \cite{deHoop60,deHoop84,deHoop85}. 

In the FDTD simulation, a computational volume of $58.2~\text{mm} \times 346.8~\text{mm} \times 58.2~\text{mm}$ was employed. Furthermore, the parameters of the configuration shown in~\cref{fig:validate} were set as $h_T = 2.1~\text{mm}$, $h_R = 2.1~\text{mm}$ and $d = 2.1~\text{mm}$, $\rho_s = 1200~\frac{\text{kg}}{\text{m}^3}$, $\lambda_s = 9.48~\text{GPa},\alpha_s = 0.0\ \frac{1}{\mu s}$, $\mu_s = 2.352~\text{GPa}$, $\rho_f =1000~\frac{\text{kg}}{\text{m}^3}$, $\lambda_f = 2.25~\text{GPa},\ \alpha_f = 0.0\ \frac{1}{\mu s}$, and $\mu_f = 0.0~\text{GPa}$.
 An linear isotropic grid size of $\Delta x = 0.15\ \text{mm}$ was employed. The thickness of the C-PML was $4.5~\text{mm}$ on all sides of the 3-D computational grid.
 The pressure values obtained at the receiver locations were sampled with a sampling rate of $40~\text{MHz}$.
 The digital representation of the line source had a Gaussian spread in the x-z-plane with a standard deviation of 1~$\text{mm}$.
The results of the FDTD simulation are superimposed on the analytical
solution in~\cref{fig:deHoop}.
The pressure profiles produced by the two methods are found to be nearly overlapping, 
indicating that the FDTD method possesses a high degree of accuracy.

In addition to validating the forward operator $\mathbb{H}$, the discrete adjoint operator was validated by application of the inner product test. The inner product test involves verifying the identity $\langle \mathbb{H}\mathbf{f},\mathbf{g}\rangle_{\mathbb{V}} = \langle \mathbf{f},\mathbb{H}^{\dagger}\mathbf{g}\rangle_{\mathbb{U}}$, where $\mathbf{f}\in \mathbb{U}$ and $\mathbf{g} \in \mathbb{V}$. Here, $\mathbb{U}$ and $\mathbb{V}$ represent the Euclidean spaces $\mathbb{R}^{N \times 1}$ and $\mathbb{R}^{LM \times 1}$, respectively. It was observed that the inner product test agreed to a 6-digit accuracy, thus validating the implementation of the discrete adjoint operator $\mathbb{H}^{\dagger}$.

\section{Computer simulation studies}
\label{sec:three}

Computer-simulation studies were conducted in which the adjoint operator
$\mathbb{H}^{\dagger}$
was employed as a heuristic  reconstruction operator.
The performance of the adjoint operator was compared with a canonical
 backprojection (BP) reconstruction algorithm~\cite{Kunyansky07,Finch04}
that assumed a homogeneous lossless fluid medium.
To further study the impact of modeling shear wave propagation in 3D transcranial PACT, additional
 computer-simulation studies were conducted to
 assess the performance of the adjoint operator for cases
 where the shear modulus of the skull was assumed to be zero. 

\subsection{Methods}
\subsubsection{Imaging geometry and phantom description}
\begin{figure}[!t]
	\centering
	\subfloat[]{\includegraphics[height=1.63in, width = 2.00in]{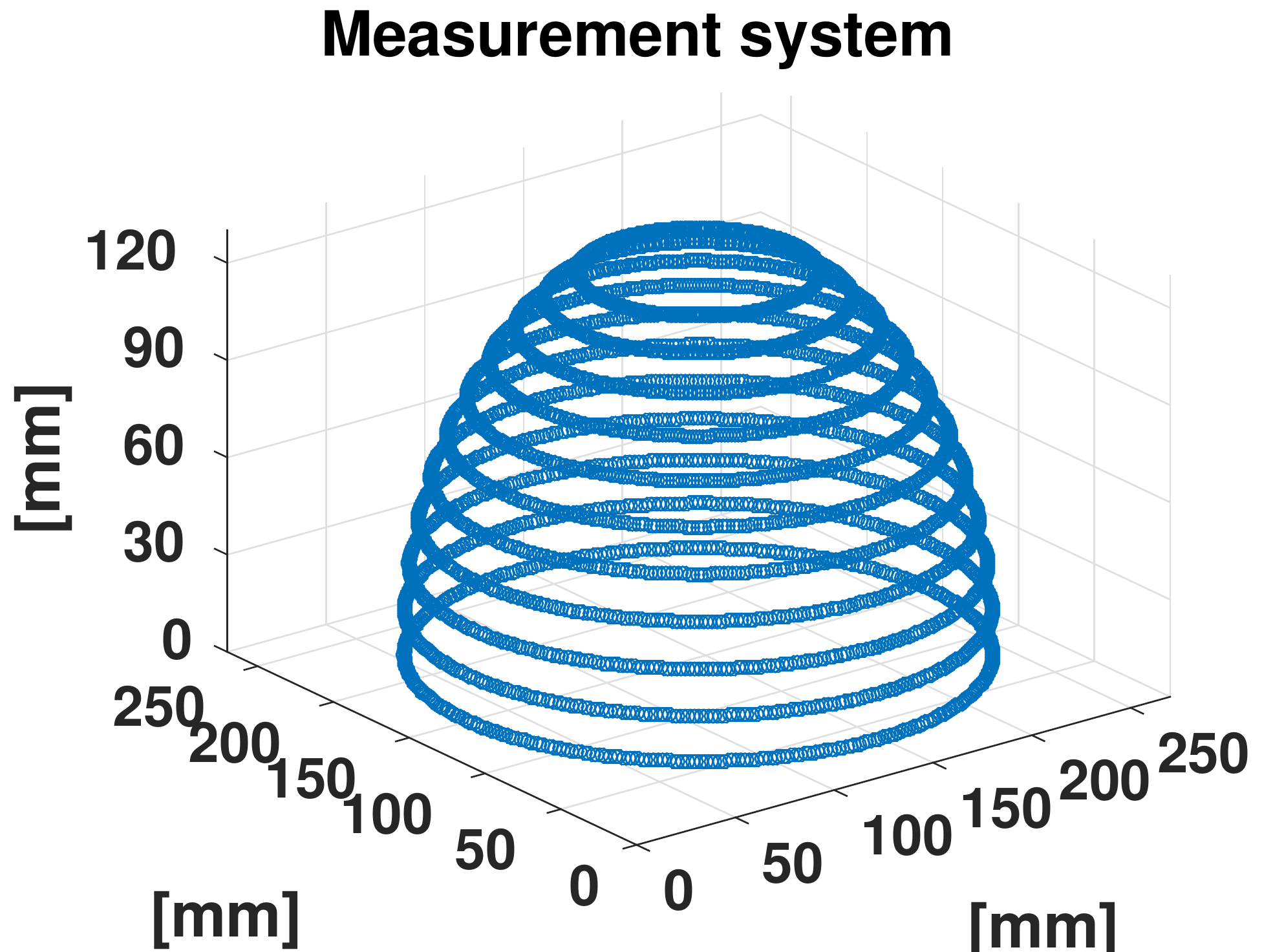}%
	\hspace{0.01in}
	\label{Fig:mp}}
	\subfloat[]{\includegraphics[height=1.63in]{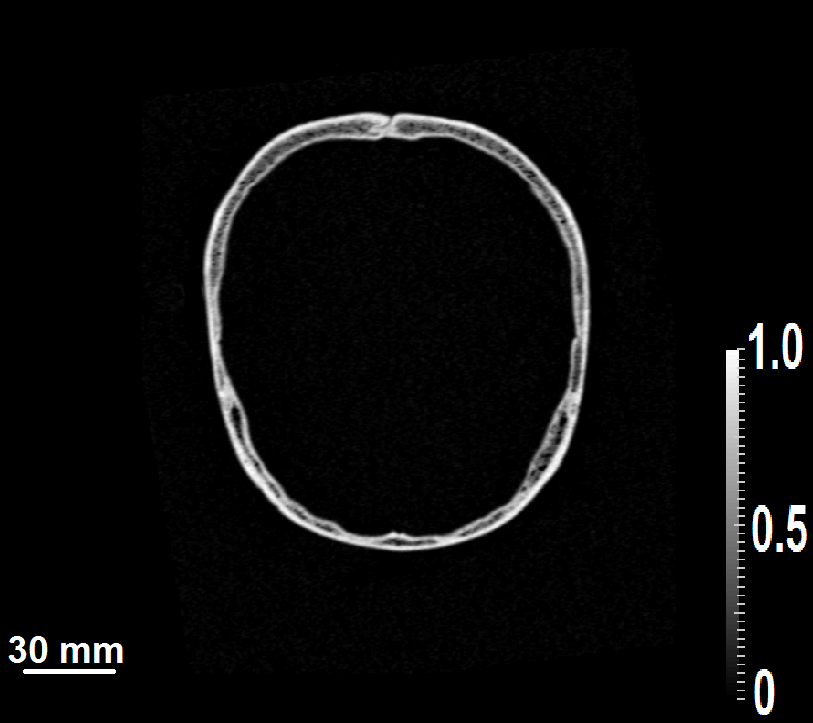}%
		\hspace{0.01in}
		\label{Fig:ctskull}}
	\subfloat[]{\includegraphics[height=1.63in]{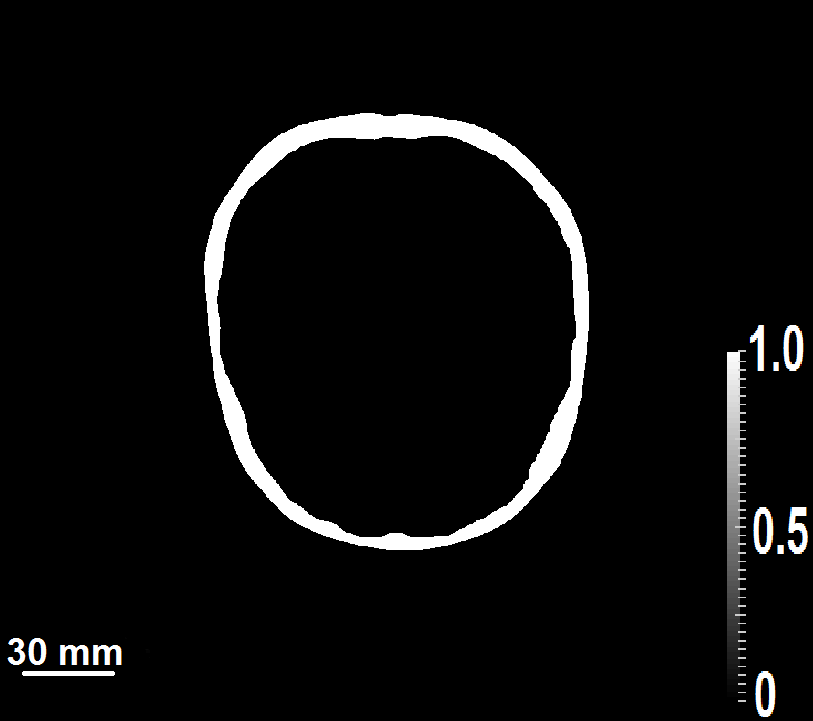}%
		\label{Fig:phiskull}}
	\caption{(a) The 3D scanning geometry used for the computer-simulation studies. (b) A 2D slice of X-ray CT image of the skull and (c) the corresponding mask generated by the segmentation algorithm. }
\end{figure}

A 3D computational volume of $270.0\ \text{mm}\times 270.0\ \text{mm}\times 135.6\ \text{mm}$ was employed.
The 3D scanning geometry, as shown in ~\cref{Fig:mp}, consisted of 11 rings of varying radii with 400 transducers  evenly distributed in each ring. 
The linear isotropic, elastic medium used in the simulation studies was generated from 3D X-ray CT images of a human skull.
 The intact human skull was purchased from Skull Unlimited International Inc. (Oklahoma City, OK) and was donated by an 83-year-old Caucasian male. The CT images were employed to infer the thickness and contour of the skull.
 
For the simulation studies involving $\mathbb{H}^{\dagger}$, we assumed the skull to be an acoustically homogeneous elastic linear isotropic medium. While we consider a relatively simple skull model, the proposed approach could also be applied for more complex skull models, such as those that consider the heterogeneity within the skull. In that case, more effort may be required to accurately estimate the acoustic properties of the skull. In order to extract the contour and location of the skull from CT images, a segmentation algorithm was employed. The segmentation algorithm generated a binary mask specifying the location of the skull within the 3D volume. A 2D slice of the CT image acquired from the human skull
 and
 the corresponding mask generated by use of the segmentation algorithm are shown in~\cref{Fig:ctskull} and~\cref{Fig:phiskull}, respectively. The medium parameters in the 3D grid were assigned such that the skull acoustic parameters ($\rho = 1850\ \frac{\text{kg}}{\text{m}^3}$, $c_l = 3.0\ \frac{\text{mm}}{\mu \text{s}}$, $c_s = 1.5 \ \frac{\text{mm}}{\mu \text{s}}$ and $\alpha = 0.1 \ \frac{1}{\mu \text{s}}$) were set at all grid positions where mask was equal to one and the background acoustic parameters ($\rho = 1000\  \frac{\text{kg}}{\text{m}^3}$, $c_l = 1.5 \ \frac{\text{mm}}{\mu \text{s}}$, $c_s = 0.0 \ \frac{\text{mm}}{\mu \text{s}}$ and $\alpha = 0.0 \ \frac{1}{\mu \text{s}}$) were set at all grid positions where mask was equal to zero. At the material interface between the skull and the background fluid medium, the density and the absorption values were arithmetically averaged to avoid any
 instability issues with the FDTD wave equation solver.

\begin{figure}[!t]
	\subfloat[]{\includegraphics[height=1.75in,width = 2.00in]{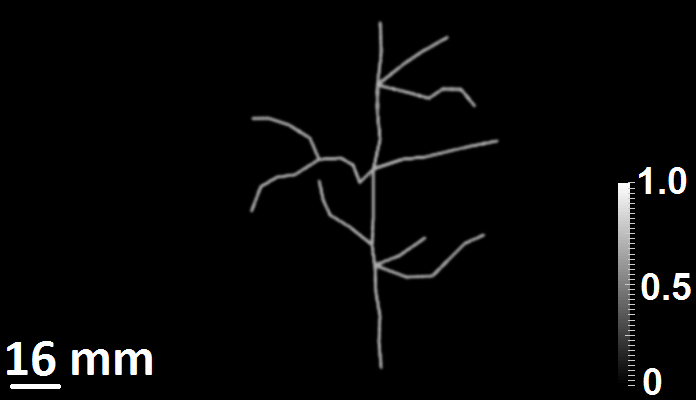}%
		\label{Fig:MAPxy2}}
	\hspace{0.01in}
	\subfloat[]{\includegraphics[height=1.75in,width = 2.00in]{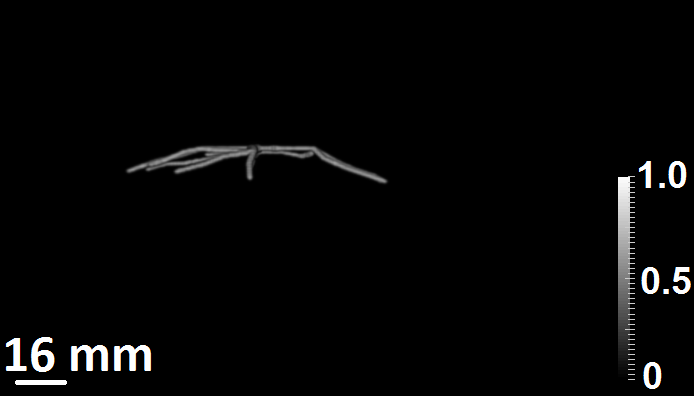}%
		\label{Fig:MAPxz2}}
	\hspace{0.01in}
	\subfloat[]{\includegraphics[height=1.75in,width = 2.00in]{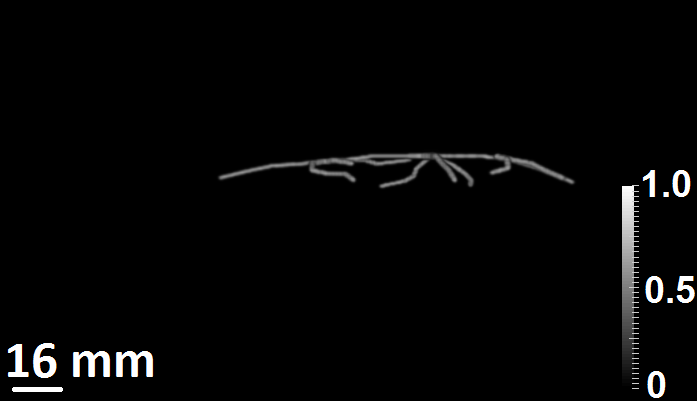}%
		\label{Fig:MAPyz2}}
	\hspace{0.01in}
	\caption{The maximum intensity projection of the initial pressure distribution  (a) along the z-axis, (b) along the y-axis, and (c) along the x-axis.}
	\label{Fig:Phantom}
\end{figure}

The initial pressure distribution assumed in the simulation studies mimicked cortical blood vessels (CBVs). The phantom, shown in~\cref{Fig:Phantom},
 consisted of CBVs positioned approximately $6\text{ mm}$ below the inner surface of the skull. The 2D maximum intensity projection images along the x-,y- and z-axis of the initial pressure distribution are shown in~\cref{Fig:Phantom}.

\subsubsection{Image reconstruction studies}

To demonstrate the use of 
$\mathbb{H}^{\dagger}$ as a reconstruction operator in 3D transcranial PACT, we conducted non-inverse crime computer-simulation studies~\cite{Colton13}, where different discretization strategies were employed to generate the measured data and to compute the action of $\mathbb{H}^{\dagger}$. In these studies, the forward data were generated using the phantom in \cref{Fig:Phantom}
 with a uniform grid size of $\Delta x = 0.225~\text{mm}$ and a temporal sampling rate of 50 MHz.
Uncorrelated Gaussian noise was added to the simulated pressure signals.
The  standard deviation of the noise was set at  5\% of the
maximum signal value recorded.
 The action of $\mathbb{H}^{\dagger}$ on the generated forward data was computed using a larger grid size of $\Delta x = 0.3~\text{mm}$. The simulated pressure data was not temporally downsampled before the application of the adjoint operator.  

Images were also reconstructed
from the noisy simulated data by use of the BP reconstruction algorithm.
 To find the optimal longitudinal SOS value for use in the BP reconstruction algorithm,
 we tuned the SOS over a range of values and picked the value that gave us the smallest mean squared error (MSE). The BP images were reconstructed using a uniform longitudinal SOS set to $1.540~\frac{mm}{\mu s}$, and using an uniform spatial grid
 of pitch $\Delta x = 0.3~\text{mm}$.

Finally, the importance of modeling shear wave propagation
in the skull was further assessed by reconstructing images by use of
a modified version of $\mathbb{H}^{\dagger}$, denoted as
$\mathbb{H}^{\dagger}_{\mu=0}$, in which the shear modulus of the elastic medium was set to zero~\cite{ChaoSkull}.
This corresponds to the un-physical situation in which the skull does not
support shear wave propagation.
 In this simulation, a uniform grid size of $\Delta x = 0.3~\text{mm}$ was used to compute the action of the adjoint operator. The longitudinal SOS of the skull
 was tuned over a range of values
and the optimal value was selected based on the MSE. Note that,
 in this case, only the longitudinal SOS of the skull
was tuned 
and the constant longitudinal SOS of the background fluid medium was 
fixed at $1.50~\frac{mm}{\mu s}$. 

\subsection{Computer-simulation: Results}
\begin{figure}[!t]
	\subfloat[]{\includegraphics[height=1.75in,width=2.0in]{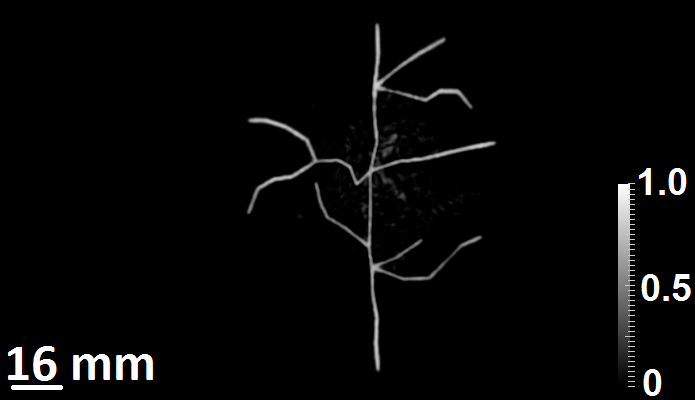}%
		\label{Fig:nonInvMAPxy}}
	\hspace{0.01in}
	\subfloat[]{\includegraphics[height=1.75in,width=2.0in]{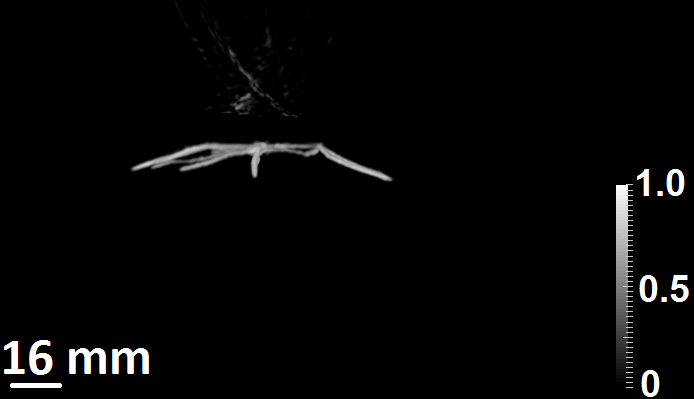}%
		\label{Fig:nonInvMAPxz}}
	\hspace{0.01in}
	\subfloat[]{\includegraphics[height=1.75in,width=2.0in]{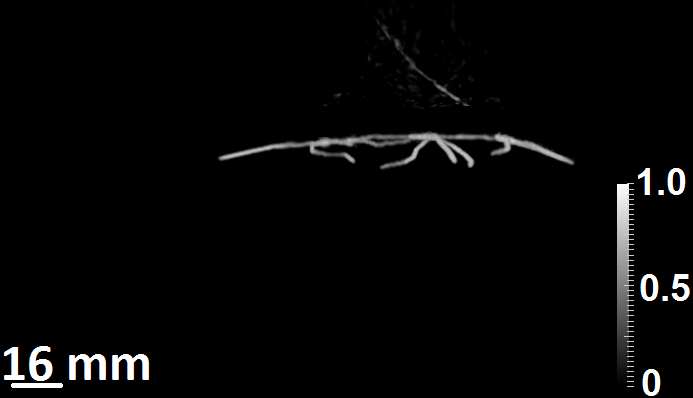}%
		\label{Fig:nonInvMAPyz}}
	\hspace{0.01in}
	\subfloat[]{\includegraphics[height=1.75in,width=2.0in]{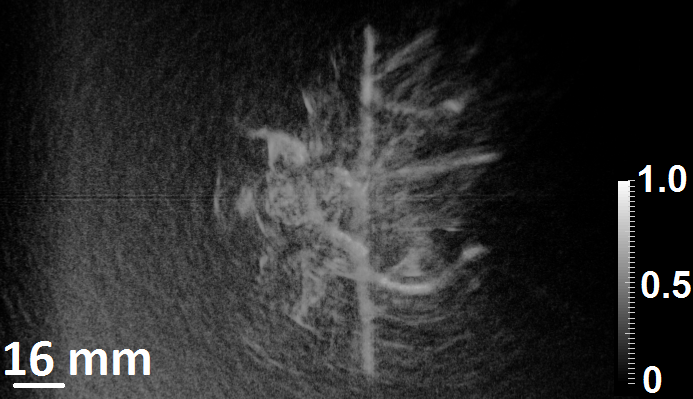}%
			\label{Fig:BPnonInvMAPxy}}
	\hspace{0.01in}
	\subfloat[]{\includegraphics[height=1.75in,width=2.0in]{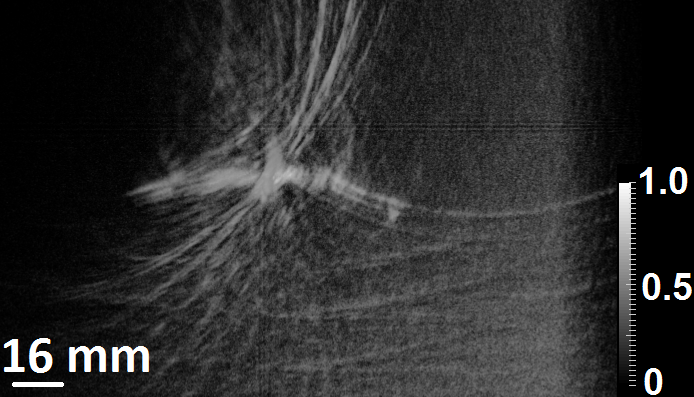}%
		\label{Fig:BPnonInvMAPxz}}
	\hspace{0.01in}
	\subfloat[]{\includegraphics[height=1.75in,width=2.0in]{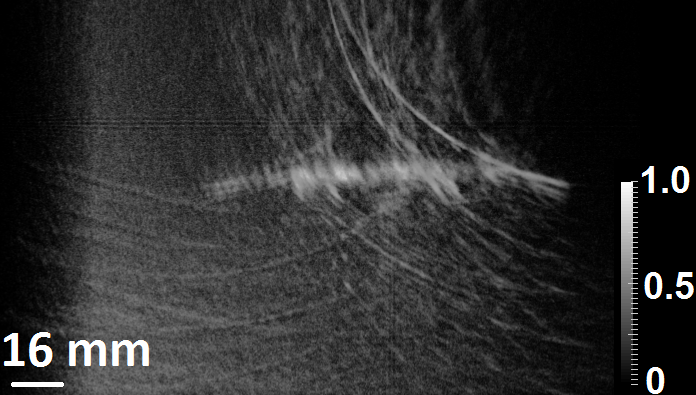}%
			\label{Fig:BPnonInvMAPyz}}
	\hspace{0.01in}
	\subfloat[]{\includegraphics[height=1.75in,width=2.0in]{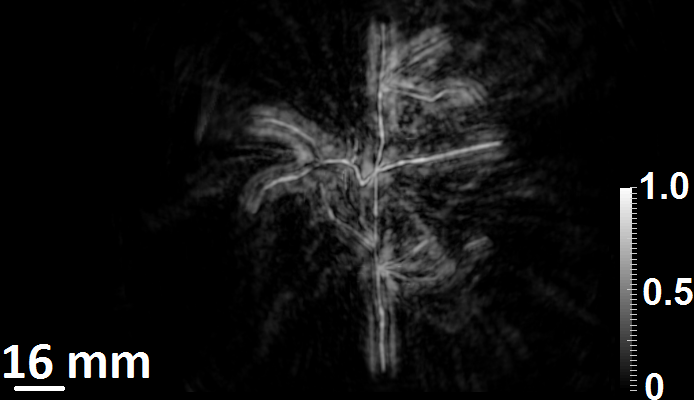}%
		\label{Fig:shearnonInvMAPxy}}
	\hspace{0.01in}
	\subfloat[]{\includegraphics[height=1.75in,width=2.0in]{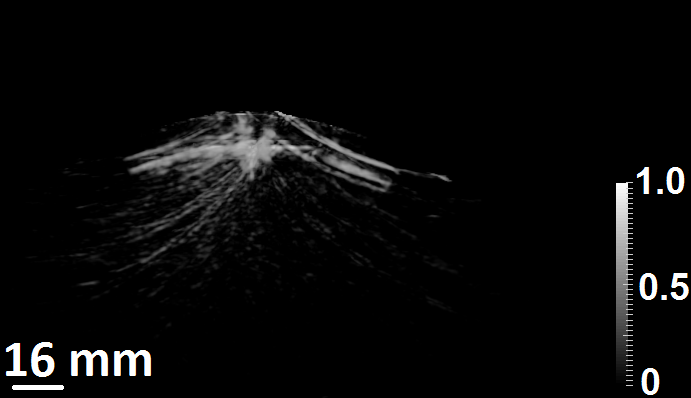}%
		\label{Fig:shearnonInvMAPxz}}
	\hspace{0.01in}
	\subfloat[]{\includegraphics[height=1.75in,width=2.0in]{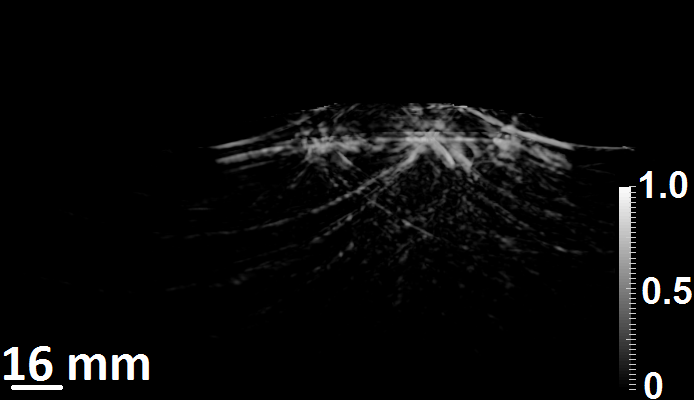}%
		\label{Fig:shearnonInvMAPyz}}
	\hspace{0.01in}
	\caption{The maximum intensity projection of the reconstructed initial pressure distribution using $\mathbb{H}^{\dagger}$, along (a) the z-axis, (b) the y-axis, and (c) the x-axis.The maximum intensity projection of the reconstructed initial pressure distribution using BP algorithm along (d) the z-axis, (e) the y-axis, and (f) the x-axis. The maximum intensity projection of the reconstructed initial pressure distribution using $\mathbb{H}_{\mu = 0}^{\dagger}$, along (g) the z-axis, (h) the y-axis, and (i) the x-axis.}
	\label{Fig:nonInv}
\end{figure}

The reconstructed images produced by use of $\mathbb{H}^{\dagger}$ 
and the BP algorithm are
 shown in~\cref{Fig:nonInvMAPxy,Fig:nonInvMAPxz,Fig:nonInvMAPyz} and~\cref{Fig:BPnonInvMAPxy,Fig:BPnonInvMAPxz,Fig:BPnonInvMAPyz}.
In both cases, the results were displayed as
 maximum intensity projection (MIP) images along three mutually perpendicular directions. 

 These results
 demonstrate that  $\mathbb{H}^{\dagger}$
 can more effectively mitigate skull-induced image distortions
 than can the BP algorithm.
Namely, despite being an approximate reconstruction operator, the image produced by application of $\mathbb{H}^{\dagger}$
accurately displays the blood vessel geometry and possesses a much cleaner background and
 contains far fewer artifacts than the image reconstructed using the BP algorithm. 
It should also be noted that in cases for which $\mathbb{H}^{\dagger}$ does
not produce images of adequate quality, a more principled iterative approach to image
reconstruction can be implemented by use of the operators $\mathbb{H}$ and
 $\mathbb{H}^{\dagger}$.


The image reconstructed by use of $\mathbb{H}^{\dagger}_{\mu=0}$ is shown
 in~\cref{Fig:shearnonInvMAPxy,Fig:shearnonInvMAPxz,Fig:shearnonInvMAPyz}.
This image contains dramatically elevated artifact levels as compared
to the one reconstructed by use of $\mathbb{H}^{\dagger}$, shown in~\cref{Fig:nonInvMAPxy,Fig:nonInvMAPxz,Fig:nonInvMAPyz}.
This demonstrates the importance of compensating for both the acoustic and the elastic
 properties of the skull in the reconstruction
 algorithm. 

\section{Experimental studies}
\label{sec:Results}
Studies that utilized experimental PACT data produced by a physical phantom were also conducted.

\subsection{Methods}
\subsubsection{Imaging geometry and phantom description}
A single element transducer was scanned over 4400 locations (11 rings with 400 evenly distributed positions per ring) in the configuration shown in~\cref{Fig:mp} to acquire the experimental data.
\begin{figure}[!t]
	    \centering
		\subfloat[]{\includegraphics[height=1.75in,width=2.5in]{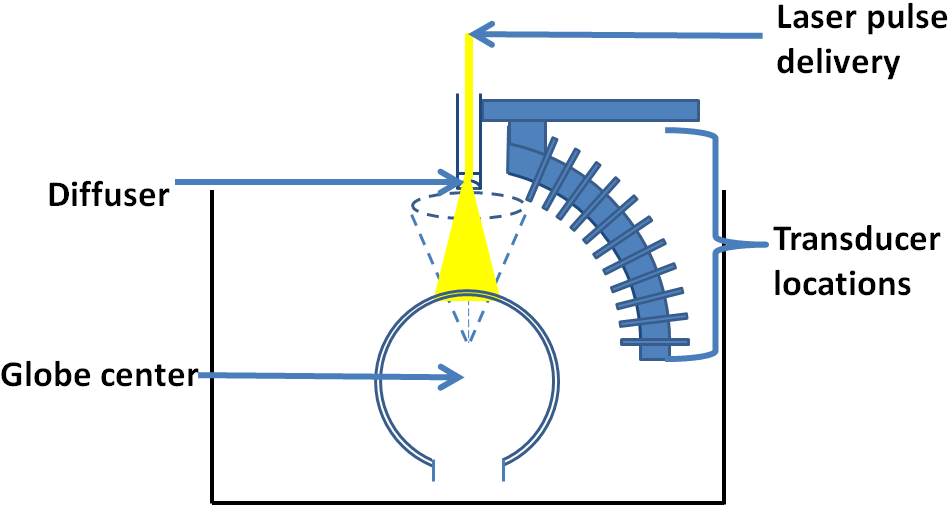}
			\label{Fig:System}}
		\subfloat[]{\includegraphics[height=1.75in,width=1.75in]{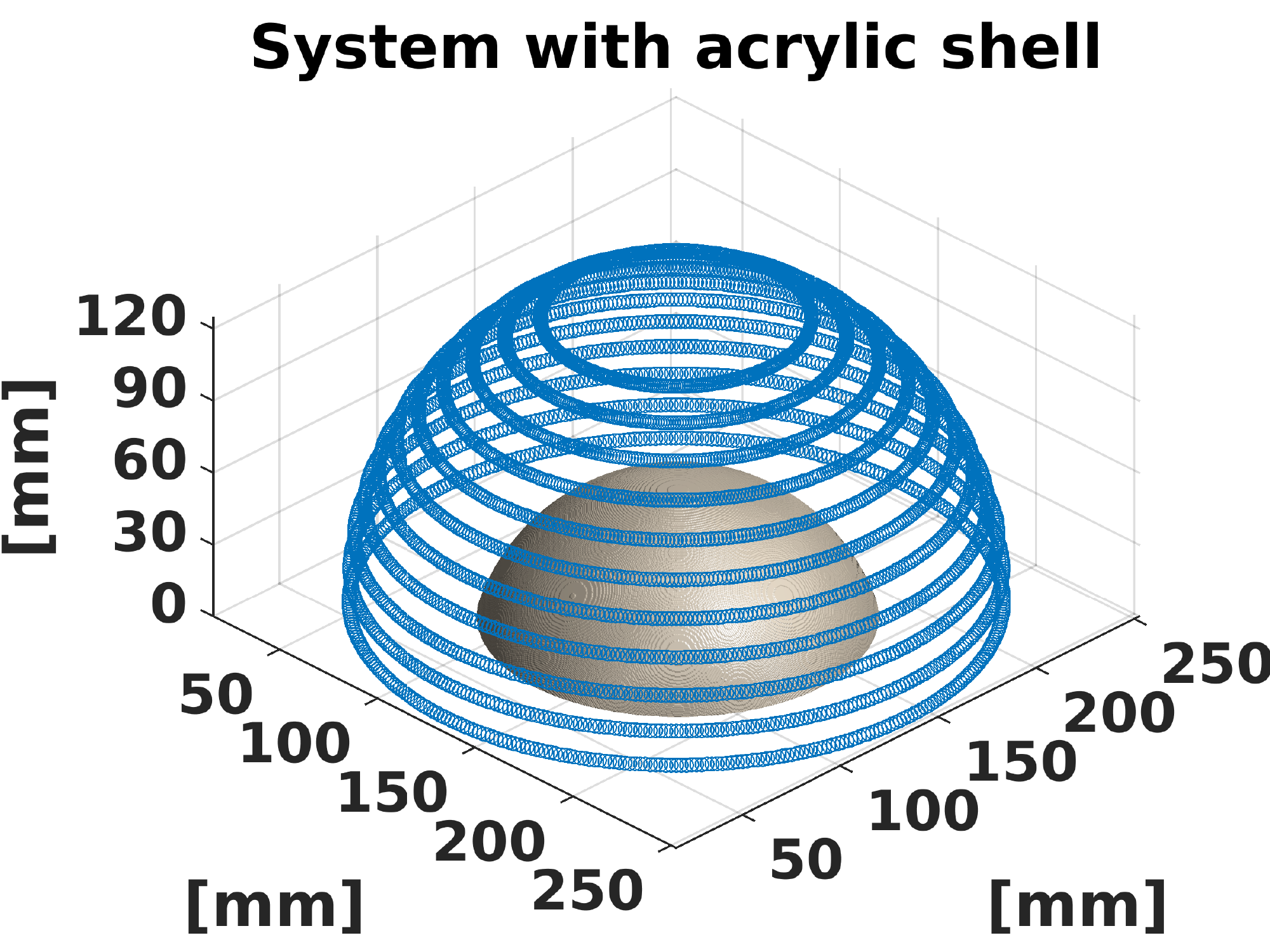}%
			\hspace{0.02in}
			\label{Fig:acrylic}}
		\subfloat[]{\includegraphics[height=1.75in,width=1.5in]{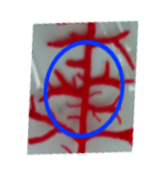}%
			\label{Fig:vessel}}
		\caption{(a) The schematic of the system. (b) The position of the acrylic globe relative to the measurement system. (c) The vessels drawn on the inner surface of the acrylic globe.}
		\label{Fig:setup}
\end{figure}
 A short 10 $ns$ laser pulse (Nd-YAG Quantel Brilliant B  laser with second harmonic generator) with a repetition rate of $10~Hz$ at a wavelength of $ 532~\text{nm}$ was used to irradiate a sample located in the center of the measurement system as shown in~\cref{Fig:System}. The subsequently generated acoustic signals were detected by unfocused transducers, with a center frequency of 1~MHz  and a bandwidth of 80~\%. The electrical signals recorded by the transducers were sampled at a temporal sampling rate of 20~MHz.

The system described above was employed to image a physical phantom
whose design was motivated by transcranial PACT.
The phantom was comprised of a  
 spherical acrylic globe of thickness of 2.5~mm
 and an inner radius of 76.2~mm, placed within a 3D volume filled with water.
The acrylic globe is an elastic solid that possesses SOS values that are
representative of a human skull.
 The location of acrylic globe relative to the
 transducer array is shown in~\cref{Fig:acrylic}.
Optically absorbing vessel-like structures were painted with Latex paint
 on the inner surface
of the acrylic globe,
 as shown in~\cref{Fig:vessel}. These vessels were intended to mimic
cortical vessels that reside near the top surface of a brain.

In order to construct $\mathbb{H}^{\dagger}$, the acoustic parameters
of the phantom need to be specified on a 3D Cartesian grid. The uniformly thick spherical acrylic shell (inner radius = 76.2~mm and thickness = 2.5~mm) was placed within the 3D volume such that the z-offset between the center of the shell and the first ring of transducer measurements was 30.2~mm. Thus, when computing $\mathbb{H}^{\dagger}$, only a spherical dome of the acrylic shell was included in the 3D simulation volume as shown by ~\cref{Fig:acrylic}.
The acoustic parameters of the 
 globe were set to be
 $\rho = 1200~\frac{\text{kg}}{\text{m}^3}$,~$c_l = 2.8~\frac{\text{m
m}}{\mu \text{s}}$,~$c_s = 1.4~\frac{\text{mm}}{\mu \text{s}}$, and $\alpha = 0.1~\frac{
1}{\mu \text{s}}$.   
 In addition, the acoustic parameters of the homogeneous background (water bath)
 were specified
 as $\rho = 1000~\frac{\text{kg}}{\text{m}^3}$,~$c_l = 1.5~\frac{\text{mm}}{\mu \text{s
}}$,~$c_s~=~0.0~\frac{\text{mm}}{\mu \text{s}}$ and $\alpha = 0.0~\frac{1}{\mu \text{s}
}$. Similar to the computer-simulation study, at the material interface between the
 globe and the background fluid medium, the density and the absorption values were
 arithmetically averaged to avoid any instability
 issues with the FDTD wave equation solver.

\subsubsection{Data preprocessing}
\begin{figure}[!t]
 	\centering
 	\subfloat[]{\includegraphics[height=2.00in, width = 3.00in]{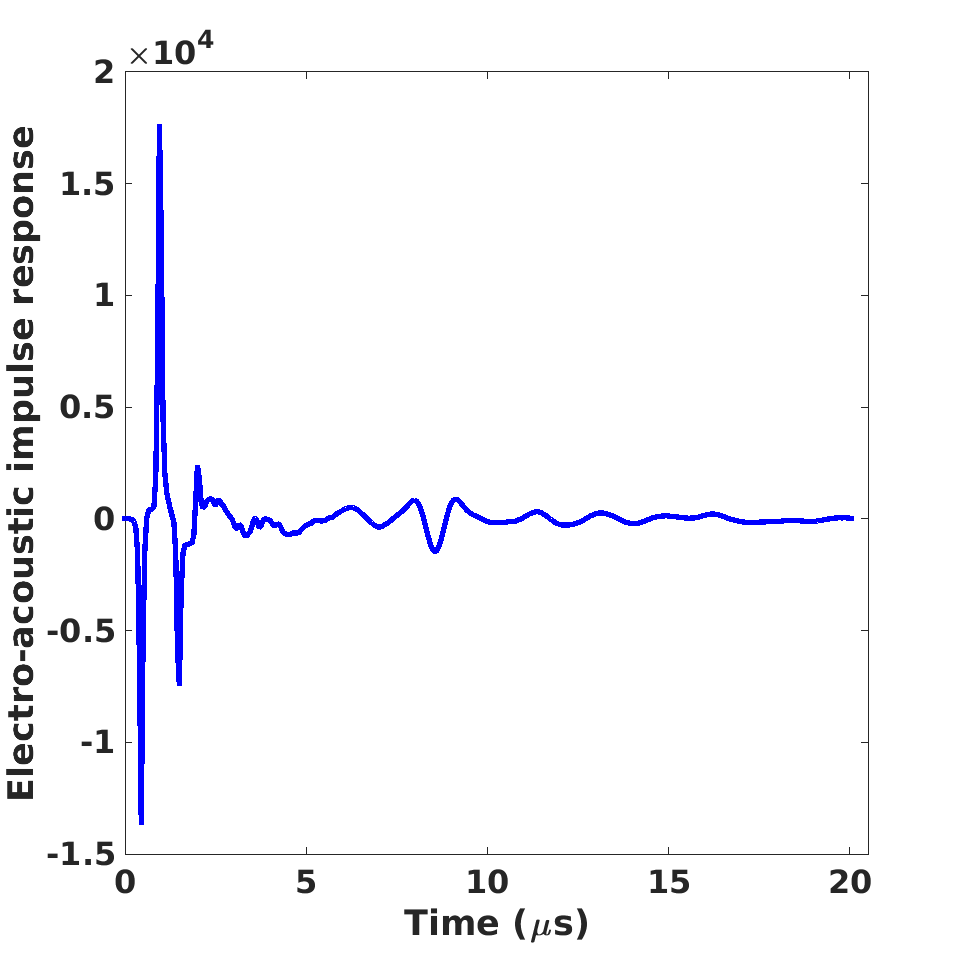}%
 		\hspace{0.01in}
 		\label{fig:EIRtime}}
 	\subfloat[]{\includegraphics[height=2.00in, width = 3.00in]{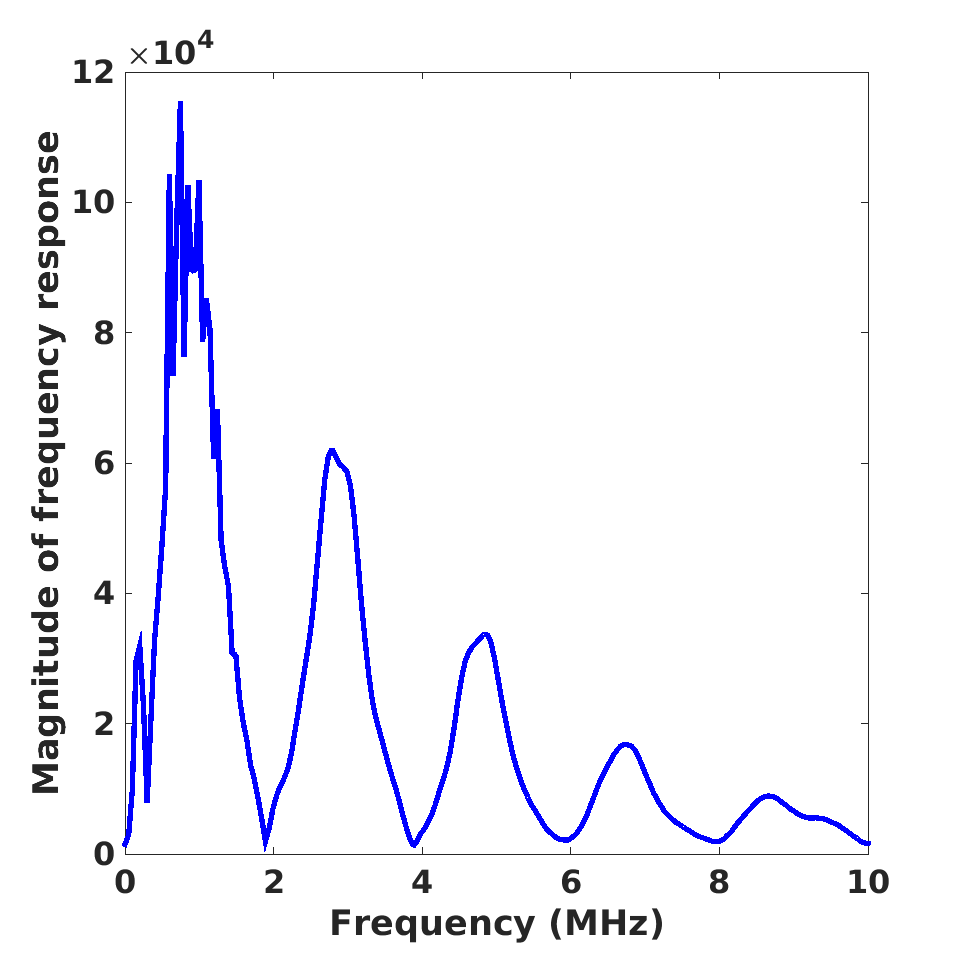}%
 		\label{fig:EIR_freq}}
 	\caption{(a) The temporal profile of the measured EIR of the transducer. (b) The magnitude of the frequency response of the EIR of the transducer.}
 	\label{Fig:EIR}
 \end{figure}

Prior to image reconstruction,
the measured data were preprocessed.
 The preprocessing involved deconvolving the acquired
 data with the electrical impulse response (EIR) of the transducer.
 The measured EIR of the transducer is shown in \cref{Fig:EIR}.
The Wiener deconvolution method was employed to extract the deconvolved PA signals from the raw electrical transducer measurements. After deconvolution, the data 
were filtered with a Hann-window low-pass filter with a cutoff frequency of 2~MHz.
The filtered data were also upsampled by a factor of 2.5, with the goal of
circumventing numerical stability issues with the wave equation solver.

\subsubsection{Image reconstruction studies}

 A 3D computational volume of
 $270~\text{mm}\times 270~\text{mm}$ $\times 135.6~\text{mm}$
 was employed in the experimental studies.
 The action of  $\mathbb{H}^{\dagger}$ on the preprocessed data was computed
 with an uniform grid size of $\Delta x = 0.3~\text{mm}$. The BP reconstruction algorithm was also applied to the preprocessed experimental data. The optimal longitudinal SOS value for the BP reconstruction algorithm was chosen by considering a range of values and selecting the value that gave us the best reconstructed image
 quality, as subjectively judged via visual inspection. For this study, the optimal longitudinal SOS of the acrylic globe was set to 1.650~$\frac{\text
 	{mm}}{\mu \text{s}}$.
 An uniform spatial grid size of $\Delta x= 0.3~\text{mm}$ was used to compute the image reconstructed using the BP algorithm. 

As in the computer-simulation studies, image were also reconstructed
by use of the operator  $\mathbb{H}^{\dagger}_{\mu=0}$.
 In this simulation, a uniform grid size of $\Delta x = 0.3~\text{mm}$ was employed.
 Moreover, the longitudinal speed of sound of the acrylic globe was tuned over a range of values. Similar to the BP algorithm, the optimal longitudinal SOS of the acrylic globe was selected based on the reconstructed image quality.

\subsection{Experimental studies: Results}
\begin{figure}[!t]
	\subfloat[]{\includegraphics[height=1.75in,width=2.0in]{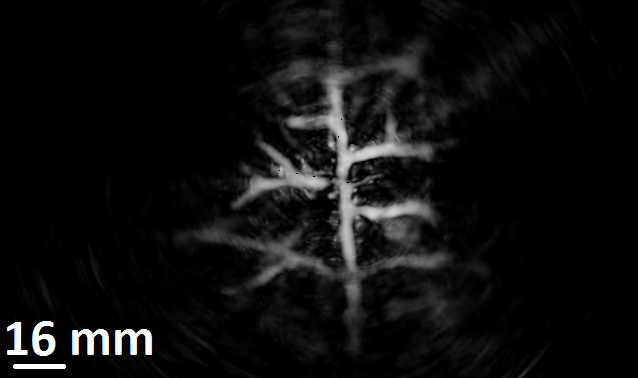}%
		\label{Fig:globeAdjMAPxy}}
	\hspace{0.01in}
	\subfloat[]{\includegraphics[height=1.75in,width=2.0in]{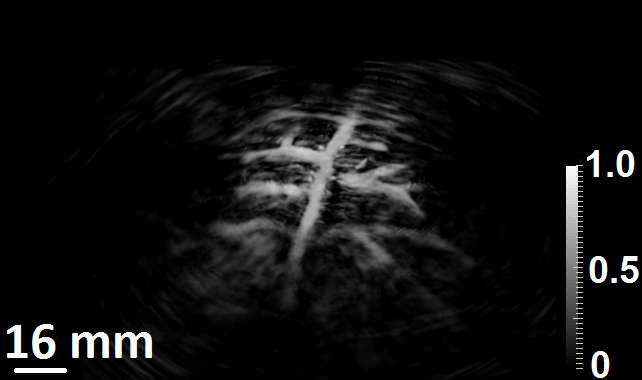}%
		\label{Fig:globeAdjMAPxz}}
	\hspace{0.01in}
	\subfloat[]{\includegraphics[height=1.75in,width=2.0in]{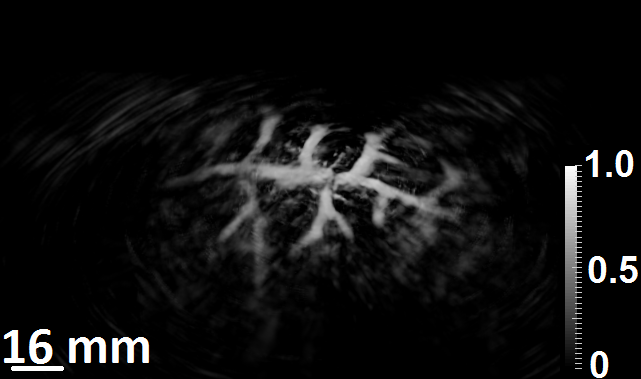}%
		\label{Fig:globeAdjMAPyz}}
	\hspace{0.01in}
	\subfloat[]{\includegraphics[height=1.75in,width=2.0in]{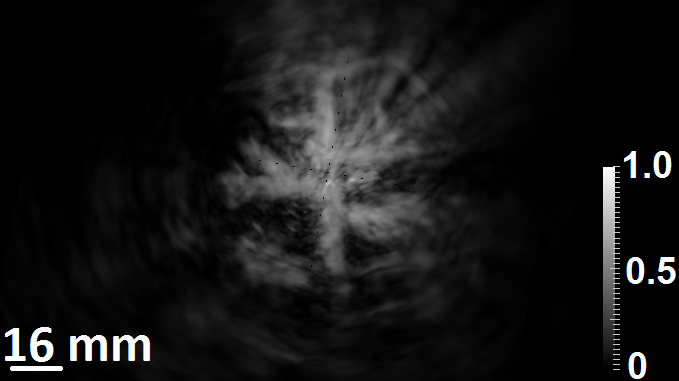}%
			\label{Fig:globeBPMAPxy}}
	\hspace{0.01in}
	\subfloat[]{\includegraphics[height=1.75in,width=2.0in]{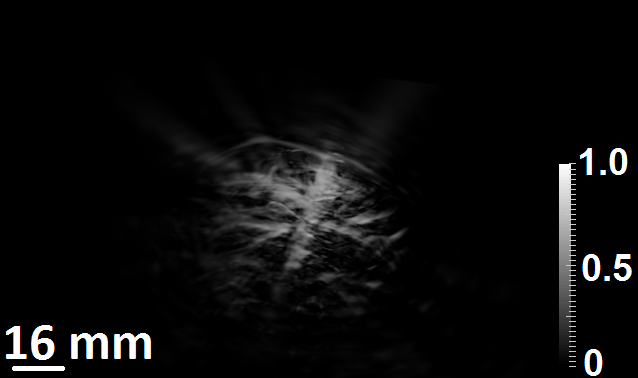}%
		\label{Fig:globeBPMAPxz}}
	\hspace{0.01in}
	\subfloat[]{\includegraphics[height=1.75in,width=2.0in]{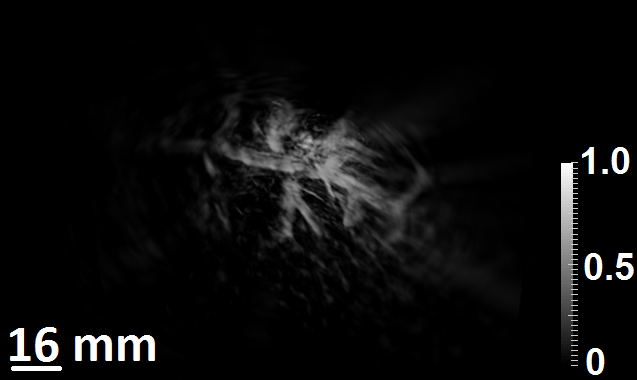}%
		\label{Fig:globeBPMAPyz}}
	\hspace{0.01in}
	\subfloat[]{\includegraphics[height=1.75in,width=2.0in]{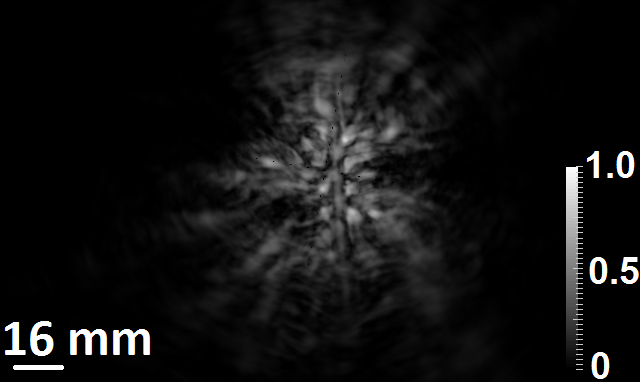}%
		\label{Fig:globeshearMAPxy}}
	\hspace{0.01in}
	\subfloat[]{\includegraphics[height=1.75in,width=2.0in]{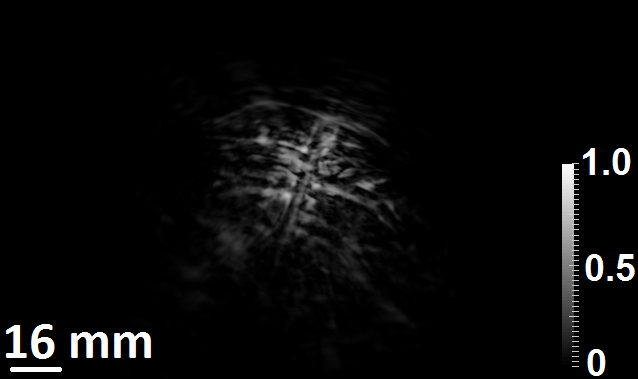}%
		\label{Fig:globeshearMAPxz}}
	\hspace{0.01in}
	\subfloat[]{\includegraphics[height=1.75in,width=2.0in]{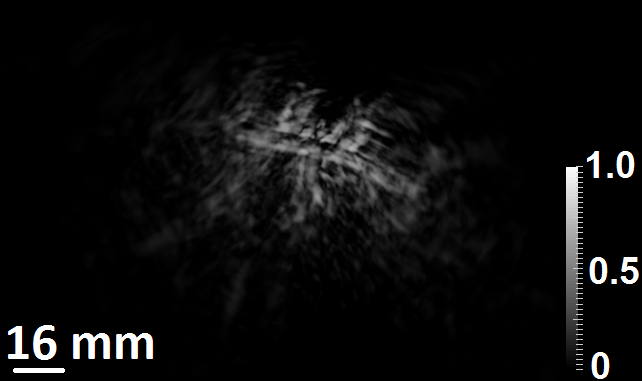}%
		\label{Fig:globeshearMAPyz}}
	\hspace{0.01in}
	\caption{The reconstructed initial pressure distribution using $\mathbb{H}^{\dagger}$ along three different views are shown in (a)-(c).The reconstructed initial pressure distribution using BP algorithm along three different views are shown in  (d)-(f).The reconstructed initial pressure distribution using $\mathbb{H}_{\mu = 0}^{\dagger}$ along three different views are shown in (g) -(i).}
	\label{Fig:globeAdj}
\end{figure}

The images reconstructed from the experimental data
 are shown in ~\cref{Fig:globeAdj}.
~\cref{Fig:globeAdjMAPxy,Fig:globeAdjMAPxz,Fig:globeAdjMAPyz} displays the image reconstructed by application of
$\mathbb{H}^{\dagger}$,
  while the image reconstructed by use of the BP algorithm is shown in
~\cref{Fig:globeBPMAPxy,Fig:globeBPMAPxz,Fig:globeBPMAPyz}. 
In the BP reconstruction algorithm, the longitudinal speed of sound of the background fluid media was set to be 1.507 $\frac{\text{mm}}{\mu \text{s}}$.  
 These results demonstrate that the $\mathbb{H}^{\dagger}$
 can more effectively mitigate image distortions due to acrylic globe (i.e., 
simulated skull structure) than can the BP algorithm.
 Additionally, some of the smaller vessel structures are not identifiable in
the BP image but are present in the image reconstructed by use of
 $\mathbb{H}^{\dagger}$.

The image reconstructed by use of  $\mathbb{H}^{\dagger}_{\mu=0}$
is shown in~\cref{Fig:globeshearMAPxy,Fig:globeshearMAPxz,Fig:globeshearMAPyz}.
This image contains dramatically elevated artifact levels as compared
to the one reconstructed by use of $\mathbb{H}^{\dagger}$, shown
in~\cref{Fig:globeAdjMAPxy,Fig:globeAdjMAPxz,Fig:globeAdjMAPyz}.
 This again demonstrates that neglecting to model the elastic properties
of the medium can lead to significant deterioration in image quality.

\section{Conclusion}

In this work, a forward and adjoint operator pair was introduced
for use in transcranial PACT.   To account for longitudinal-to-shear-wave
conversions within the skull, these operators were based on a
3D elastic wave equation.
 Massively parallel implementations
 of these operators employing multiple graphics
 processing units (GPUs) were also developed.
The developed numerical framework was validated and investigated in computer-simulation
and experimental phantom studies whose designs were
 motivated by transcranial PACT applications.

The numerical studies presented employed the 
adjoint operator $\mathbb{H}^{\dagger}$ and a canonical BP
reconstruction operator for image reconstruction. The two reconstruction approaches differ in that
 the operator $\mathbb{H}^{\dagger}$ is based on the elastic wave
equation for a specified heterogeneous elastic medium, while the canonical BP
method assumes a homogeneous fluid medium.
Our results demonstrated
 that neglecting to model the heterogeneous elastic properties
of the medium can lead to significant deterioration in image quality.
Although not presented, use of a filtered BP algorithm that was 
formed by discretizing an exact inversion formula \cite{Finch04,Xu05}
 did not alter the presented findings. In addition, the numerical and experimental results also demonstrated that the failure to model the shear wave propagation in elastic isotropic media can lead to significant deterioration in image quality.

There remain several important topics for further investigation. The proposed attenuation model is a diffusive model, where the attenuation coefficient is independent of the frequency. However, attenuation of elastic waves propagating in the skull is frequency-dependent. Therefore,
 the accuracy of proposed matched operator pair can 
 be improved by incorporating a frequency dependent attenuation model. Furthermore, in both the computer-simulation and the experimental studies, to perform accurate reconstruction we require prior knowledge of the spatial distribution of the elastic parameters of the medium.  Estimating the elastic parameters in an human skull is a challenging task, as the human skull is an acoustically heterogenous medium. The skull is a three-layer structure with a porous zone, called the diploe, stacked between two dense layers, the outer and inner tables~\cite{aubry2003experimental}. Recently, the possibility to deduce acoustic properties of the skull from adjunct CT data has gained some traction~\cite{Fink,jones2013transcranial,aubry2003experimental,connor02}. As CT images provide information about the internal structure of the skull, it has been used to estimate the internal heterogeneities in density, speed and absorption. Hence, a study investigating the impact of errors in estimating acoustic parameters using adjunct CT data on the
 accuracy of the reconstructed image can be a subject of further study.

 In addition, the use of  $\mathbb{H}$ and $\mathbb{H}^{\dagger}$
in studies of iterative image reconstruction is a natural topic for
investigation.
 The developed massively parallel implementations of
 $\mathbb{H}$ and $\mathbb{H}^{\dagger}$
 using multiple GPUs will facilitate these studies.
 For example, the matched operator pair can be employed in an
 iterative image reconstruction method
 that seeks to minimize a penalized least square functional.
Image reconstruction approaches such as this will be necessary 
when $\mathbb{H}^{\dagger}$ does not produce useful images.

\section*{Appendix- Convolutional PML (C-PML)}
\subsection{Introduction to C-PML}
One of the most popular methods of implementing absorbing boundary
 conditions is the PML. It was originally developed by
 Berenger for use with Maxwell's equations using a split-field formulation~\cite{Berenger94,Berenger96}. One of the drawbacks to the split-field formulation of the classical PML is that it increases the number of field variables that need to be stored and the number of differential equations that need to be solved over the whole domain.

Later, a complex coordinate stretching technique was developed~\cite{Roden00,Chew94}, which is equivalent to Berenger's formulation in the frequency-domain, but which has a much more efficient implementation in the time-domain. In this approach, the spatial derivatives along each direction are replaced with a scaled version. For example, in the $x$-direction, the scaled spatial derivative is given by
	\begin{align}
	\partial_{\tilde{x}} = \frac{1}{S_x\left(x, \omega\right)} \partial_x \label{eqn:mod_deriv},
	\end{align}
	where $S_x\left(x,\omega\right)$ is a spatially-dependent, complex scaling factor given in the temporal frequency domain by
	\begin{align}
	S_x\left(x,\omega\right) = 1 + \frac{d_x\left(x\right)}{j\omega}.
	\end{align}
	Here, $d_x(x)$ is the profile for the absorption of the absorbing boundary layer, which is equal to zero outside of the layer. In the time-domain, Eqn.~\cref{eqn:mod_deriv} is given by
	\begin{align}\label{eq:smalls}
	\partial_{\tilde{x}} = s_x\left(x,t\right) *_t \partial_x ,
	\end{align}
	which is why this method is also sometimes called a convolutional PML (C-PML).
	
	While Berenger's formulation of the PML gives rise to an absorbing boundary layer with a reflection coefficient of zero for all angles of incidence, this property only holds in the continuous case. Upon discretization, for waves arriving at grazing incidence, a large amount of energy is sent back into the main domain in the form of spurious reflected waves. This makes the discrete classical PML less efficient for cases where the sources are located close to the edge of the grid, which is commonly encountered in elastic wave modeling~\cite{Komatitsch07}. To circumvent this problem, a more general form for the scaling factor was developed~\cite{Kuzuoglu99,Roden00}, given by
	\begin{align}\label{eq:SforPML}
	S_x\left(x, \omega\right) = \kappa_x\left(x\right) + \frac{d_x\left(x\right)}{\beta_x\left(x\right) + j\omega}.
	\end{align}
	Given Eqns.~\cref{eq:SforPML,eqn:mod_deriv,eq:smalls},  one can write 
	\begin{align}
	s_x\left(x, t\right) = \frac{1}{\kappa_x\left(x\right)} \delta\left(t\right) + \zeta_x\left(x,t\right) ,
	\end{align}
	where
	\begin{align}
	\zeta_x\left(x,t\right) \equiv \frac{d_x\left(x\right)}{\kappa_x\left(x\right)^2} \exp\left(- \left[d_x\left(x\right)/\kappa_x\left(x\right) + \beta_x\left(x\right)\right] t\right) U(t),
	\end{align}
	and $U(t)$ denotes the Heaviside step function. Thus, our modified spatial derivative operator is given by
	\begin{align}
	\partial_{\tilde{x}} = \frac{1}{\kappa_x\left(x\right)} \partial_x + \zeta_x\left(t\right) *_t \partial_x.
	\end{align}
	From a numerical point of view, the calculation of the convolution is costly because it requires at each time step to sum over all previous time steps. However, given the particular form of $\zeta_i$, it is possible to compute the convolution using a recursive technique by use of a memory variable~\cite{Luebbers92}. Given a memory variable $\phi_x$, the derivative of a field variable along the $x^{th}$ direction can be written as
	\begin{align}\label{eq:dCPML}
	\partial_{\tilde{x}} = \frac{1}{\kappa_x} \partial_x + \phi_x,
	\end{align}
	where this memory variable is updated recursively in time according to
	\begin{align}\label{eq:phiPML}
	\phi_x^n = b_x \phi_x^{n-1} + a_x \left(\partial_x\right)^{n-\frac{1}{2}} ,
	\end{align}
	where the spatial derivative acts on the specific field variable associated with the memory variable $\phi_x(x)$. In addition, the PML decay constants $a_x(x)$ and $b_x(x)$ in Eqn.~\cref{eq:phiPML} are given by 
	\begin{subequations}
		\begin{align}
		b_x\left(x\right) &= \exp\left(-\left(d_x\left(x\right)/\kappa_x\left(x\right) +\beta_x\left(x\right)\right)\Delta t\right) \\
		a_x\left(x\right) &= \frac{d_x}{\kappa_x \left(d_x + \kappa_x \beta_x\left(x\right)\right)}\left(b_x\left(x\right) - 1\right) .
		\end{align}
	\end{subequations}
	\subsection{Application of C-PML to elastic wave equation}
	In order to apply the C-PML-based absorbing boundary conditions to the elastic wave equation in Eqn.~\cref{eq:Elastic}, one can replace the spatial derivative operations defined in Eqn.~\cref{eq:Elastic}, with the modified spatial derivative operations defined in Eqn.~\cref{eq:dCPML}. For simplicity, we consider $\kappa = 1$. The parameter $\kappa$ was introduced to attenuate evanescent waves when modeling Maxwell's equations and may be less critical to elastic wave simulations \cite{Komatitsch07}. Thus, given the modified spatial derivative operators in Eqn.~\cref{eq:dCPML}, the collection of differential equations defined in  Eqn.~\cref{eq:Elastic} can be written as
	\begin{subequations}\label{eq:collaux}
		\begin{align}
		\partial_t \dot{u}^1 + \alpha \dot{u}^1 &= \frac{1}{\rho} \sum_{i=1}^{3} \left(\partial_i \sigma^{1i} + \zeta_i * \partial_i \sigma^{1i} \right), \\
		\partial_t \dot{u}^2 + \alpha \dot{u}^2 &= \frac{1}{\rho} \sum_{i=1}^{3} \left( \partial_i \sigma^{2i} + \zeta_i * \partial_i \sigma^{2i}\right), \\
		\partial_t \dot{u}^3 + \alpha \dot{u}^3 &= \frac{1}{\rho} \sum_{i=1}^{3} \left( \partial_i \sigma^{3i} + \zeta_i * \partial_i \sigma^{3i}\right), \\
		\partial_t \sigma^{11} &= \lambda \sum_{i=1}^3 \left(\partial_i u^i + \zeta_i * \partial_i u^i\right) + 2\mu \left( \partial_1 \dot{u}^1 + \zeta_1 * \partial_1 \dot{u}^1 \right), \\
		\partial_t \sigma^{22} &= \lambda \sum_{i=1}^3 \left(\partial_i u^i  + \zeta_i * \partial_i u^i\right) + 2\mu \left( \partial_2 \dot{u}^2 + \zeta_2 * \partial_2 \dot{u}^2 \right), \\
		\partial_t \sigma^{33} &= \lambda \sum_{i=1}^3 \left(\partial_i u^i  + \zeta_i * \partial_i u^i\right) + 2\mu \left( \partial_3 \dot{u}^3 + \zeta_3 * \partial_3 \dot{u}^3 \right), \\
		\partial_t \sigma^{23} &= \mu \left( \partial_2 \dot{u}^3 + \partial_3 \dot{u}^2 + \zeta_2 * \partial_2 \dot{u}^3 + \zeta_3 * \partial_3 \dot{u}^2 \right), \\
		\partial_t \sigma^{13} &= \mu \left( \partial_1 \dot{u}^3 + \partial_3 \dot{u}^1 + \zeta_1 * \partial_1 \dot{u}^3 + \zeta_3 * \partial_3 \dot{u}^1 \right) \text{ and }\\ 
		\partial_t \sigma^{12} &= \mu \left( \partial_1 \dot{u}^2 + \partial_2 \dot{u}^1 + \zeta_1 * \partial_1 \dot{u}^2 + \zeta_2 * \partial_2 \dot{u}^1 \right).
		\end{align}
	\end{subequations}
	Note that, because the elastic media is linear isotropic ($\sigma^{ij} = \sigma^{ji}$), it is not necessary to compute all nine components of the stress tensor. Let the memory variables be defined as
	\begin{subequations}\label{eq:Memvar}
		\begin{align}
		\phi^{ij} &\gets \zeta_j * \partial_j \sigma^{ij} \\
		\psi^{ij} &\gets \zeta_j * \partial_j u^i,
		\end{align}
	\end{subequations}
	where the memory variables are updated according to
	\begin{subequations}
		\begin{align}
		\phi^{ij}_{m + \frac{1}{2}} &= b_j\left(r^{ij}\right) \phi_{m-\frac{1}{2}}^{ij} + a_j\left(r^{ij}\right) \partial_j \sigma^{ij}_{m} \\
		\psi^{ij}_{m} &= b_j\left(r^i\right) \psi^{ij}_{m - 1} + a_j\left(r^i\right) \partial_j u^i_{m -\frac{1}{2}}.
		\end{align}
	\end{subequations}
	Substituting the memory variables defined in Eqn.~\cref{eq:Memvar} into Eqn.~\cref{eq:collaux} we have
	\begin{subequations}\label{eq:collaux2}
		\begin{align}
		\partial_t \dot{u}^1 + \alpha \dot{u}^1 &= \frac{1}{\rho} \sum_{i=1}^{3} \left(\partial_i \sigma^{1i} + \phi^{1i} \right), \\
		\partial_t \dot{u}^2 + \alpha \dot{u}^2 &= \frac{1}{\rho} \sum_{i=1}^{3} \left( \partial_i \sigma^{2i} + \phi^{2i}\right), \\
		\partial_t \dot{u}^3 + \alpha \dot{u}^3 &= \frac{1}{\rho} \sum_{i=1}^{3} \left( \partial_i \sigma^{3i} + \phi^{3i}\right), \\
		\partial_t \sigma^{11} &= \lambda \sum_{i=1}^3 \left(\partial_i u^i + \psi^{ii}\right) + 2\mu \left( \partial_1 \dot{u}^1 + \psi^{11} \right), \\
		\partial_t \sigma^{22} &= \lambda \sum_{i=1}^3 \left(\partial_i u^i  + \psi^{ii}\right) + 2\mu \left( \partial_2 \dot{u}^2 + \psi^{22} \right), \\
		\partial_t \sigma^{33} &= \lambda \sum_{i=1}^3 \left(\partial_i u^i  + \psi^{ii}\right) + 2\mu \left( \partial_3 \dot{u}^3 + \psi^{33} \right), \\
		\partial_t \sigma^{23} &= \mu \left( \partial_2 \dot{u}^3 + \partial_3 \dot{u}^2 + \psi^{23} + \psi^{32} \right), \\
		\partial_t \sigma^{13} &= \mu \left( \partial_1 \dot{u}^3 + \partial_3 \dot{u}^1 + \psi^{13} + \psi^{31} \right), \text{ and } \\
		\partial_t \sigma^{12} &= \mu \left( \partial_1 \dot{u}^2 + \partial_2 \dot{u}^1 + \psi^{12} + \psi^{21} \right).
		\end{align}
	\end{subequations}	
\subsection{Discrete formulation}	
In order to explicitly write the forward and the adjoint operator for the FDTD elastic wave equation solver with the C-PML, we need to define the following matrices:
	\begin{subequations}\label{eq:constPML}
		\begin{align}
		\boldsymbol{\mathcal{A}}_{ij} &\equiv \text{diag}\left(a_i\left(\mathbf{r}_1^{j}\right), \cdots, a_i\left(\mathbf{r}_N^j\right)\right)\\
		\widetilde{\boldsymbol{\mathcal{A}}}_{ij} &\equiv \text{diag}\left(a_i\left(\mathbf{r}_1^{ij}\right), \cdots, a_i\left(\mathbf{r}_N^{ij}\right)\right) \\
		\boldsymbol{\mathcal{B}}_{ij} &\equiv \text{diag}\left(b_i\left(\mathbf{r}_1^{j}\right), \cdots, b_i\left(\mathbf{r}_N^j\right)\right) \\
		\widetilde{\boldsymbol{\mathcal{B}}}_{ij} &\equiv \text{diag}\left(b_i\left(\mathbf{r}_1^{ij}\right), \cdots, b_i\left(\mathbf{r}_N^{ij}\right)\right) \\
		\boldsymbol{\mathcal{X}}_{ij} &\equiv \text{diag}\left(\chi^i\left(\mathbf{r}_1^{j}\right), \cdots, \chi^i\left(\mathbf{r}_N^j\right)\right) \\
		\widetilde{\boldsymbol{\mathcal{X}}}_{ij} &\equiv \text{diag}\left(\chi^i\left(\mathbf{r}_1^{ij}\right), \cdots, \chi^i\left(\mathbf{r}_N^{ij}\right)\right),
		\end{align}
	\end{subequations}
	where $a_i\left(\mathbf{r}\right), b_i\left(\mathbf{r}\right)$ are PML decay constants in the $i^{th}$ direction and $\chi^i\left(\mathbf{r}\right)$ represents the PML indicator function which is one if the component of $\mathbf{r}$ along the $i^{th}$ direction is in the PML.
	Let the PML memory variable for the spatial derivative of $\boldsymbol{\sigma}^{ij}$ along the $j$-th direction be denoted as
 $\boldsymbol{\phi}^{ij}$, and the PML memory variable for the spatial derivative of $\dot{\mathbf{u}}^i$ along the $j$-the direction be denoted as $\boldsymbol{\psi}^{ij}$. It should be noted that $\boldsymbol{\phi}^{ij}$ is sampled at $\mathbf{r}^j$ and $\boldsymbol{\psi}^{ij}$ at $\mathbf{r}^{ij}$. Let us define
	\begin{subequations}\label{eq:auxPML}
		\begin{align}
		\overline{\boldsymbol{\phi}}_m &\equiv \left[\boldsymbol{\phi}^{11}_m, \boldsymbol{\phi}^{12}_m, \boldsymbol{\phi}^{13}_m, \boldsymbol{\phi}^{21}_m, \boldsymbol{\phi}^{22}_m, \boldsymbol{\phi}^{23}_m, \boldsymbol{\phi}^{31}_m, \boldsymbol{\phi}^{32}_m, \boldsymbol{\phi}^{33}_m \right]^T \\
		\overline{\boldsymbol{\psi}}_m &\equiv \left[\boldsymbol{\psi}^{11}_m, \boldsymbol{\psi}^{12}_m, \boldsymbol{\psi}^{13}_m, \boldsymbol{\psi}^{21}_m,  \boldsymbol{\psi}^{22}_m, \boldsymbol{\psi}^{23}_m, \boldsymbol{\psi}^{31}_m, \boldsymbol{\psi}^{32}_m, \boldsymbol{\psi}^{33}_m\right]^T,
		\end{align}
	\end{subequations}
	where $\overline{\boldsymbol{\phi}}_m \in \mathbb{R}^{9N \times 1}$ and $\overline{\boldsymbol{\psi}}_m \in \mathbb{R}^{9N \times 1}$.
	
	Further, let us define the operators
	\begin{subequations}\label{eq:opPML}
		\begin{align}
		\mathbf{B}_{ij} \overline{\boldsymbol{\phi}}_m &\equiv \boldsymbol{\mathcal{B}}_{ij} \boldsymbol{\phi}_m^{ij} ,\\
		\widetilde{\mathbf{B}}_{ij} \overline{\boldsymbol{\psi}}_m &\equiv \widetilde{\boldsymbol{\mathcal{B}}}_{ij} \boldsymbol{\psi}_m^{ij} ,\\
		\mathbf{E}_{ij} \overline{\boldsymbol{\sigma}}_m &\equiv \boldsymbol{\mathcal{A}}_{ij} \partial_j \boldsymbol{\sigma}^{ij}_m,\\
		\widetilde{\mathbf{E}}_{ij} \overline{\dot{\mathbf{u}}}_m &\equiv \widetilde{\boldsymbol{\mathcal{A}}}_{ij} \partial_j \dot{\mathbf{u}}_m^i ,\\
		\boldsymbol{\Gamma}_{i} \overline{\boldsymbol{\phi}}_m &\equiv \Delta t \mathbf{Q}_i^{-1} \sum_{j=1}^{3} \boldsymbol{\mathcal{X}}_{ij} \boldsymbol{\phi}_m^{ij} ,\text{ and }\\
		\boldsymbol{\Delta}_{ij} \overline{\boldsymbol{\psi}}_m &\equiv \Delta t \left[\delta_{ij} \boldsymbol{\Lambda}_{ij} \sum_{k=1}^{3} \boldsymbol{\psi}_m^{kk} + \boldsymbol{\mathcal{M}}_{ij} \left(\widetilde{\boldsymbol{\mathcal{X}}}_{ij} \boldsymbol{\psi}_m^{ij} + \widetilde{\boldsymbol{\mathcal{X}}}_{ji} \boldsymbol{\psi}_m^{ji}\right)\right],
		\end{align}
	\end{subequations}
	where $\mathbf{B}_{ij}, \widetilde{\mathbf{B}}_{ij}, \boldsymbol{\Gamma}_i, \boldsymbol{\Delta}_{ij} \in \mathbb{R}^{N \times 9N}$, $\mathbf{E}_{ij} \in \mathbb{R}^{N \times 6N}$, and $\widetilde{\mathbf{E}}_{ij} \in \mathbb{R}^{N \times 3N}$.
	Given the definitions in Eqns. \cref{eq:constPML,eq:auxPML,eq:opPML}, we can write the discretized form of Eqn.~\cref{eq:collaux2} as 
	\begin{subequations}
		\begin{align}
		\boldsymbol{\phi}_{m + \frac{1}{2}}^{ij} &= \mathbf{B}_{ij} \overline{\boldsymbol{\phi}}_{m-\frac{1}{2}} + \mathbf{E}_{ij} \overline{\boldsymbol{\sigma}}_m \\
		\dot{\mathbf{u}}_{m+\frac{1}{2}}^i &= \mathbf{J}_i \overline{\dot{\mathbf{u}}}_{m-\frac{1}{2}} + \boldsymbol{\Phi}_i \overline{\boldsymbol{\sigma}}_m + \boldsymbol{\Gamma}_i \overline{\boldsymbol{\phi}}_{m + \frac{1}{2}} \\
		\boldsymbol{\psi}_{m+1}^{ij} &= \widetilde{\mathbf{B}}_{ij} \overline{\boldsymbol{\psi}}_{m} + \widetilde{\mathbf{E}}_{ij} \overline{\dot{\mathbf{u}}}_{m+\frac{1}{2}} \\
		\boldsymbol{\sigma}_{m+1}^{ij} &= \boldsymbol{\sigma}_{m}^{ij} + \boldsymbol{\Psi}_{ij} \overline{\dot{\mathbf{u}}}_{m+\frac{1}{2}} + \boldsymbol{\Delta}_{ij} \overline{\boldsymbol{\psi}}_{m+1}.
		\end{align}
	\end{subequations}
	Further, let us also define
	\begin{subequations}
		\begin{align}
		\mathbf{B} &\equiv \left[\mathbf{B}_{11}, \mathbf{B}_{12}, \mathbf{B}_{13}, \mathbf{B}_{21}, \mathbf{B}_{22}, \mathbf{B}_{23}, \mathbf{B}_{31}, \mathbf{B}_{32}, \mathbf{B}_{33} \right]^T ,\\
		\widetilde{\mathbf{B}} &\equiv \left[\widetilde{\mathbf{B}}_{11}, \widetilde{\mathbf{B}}_{12}, \widetilde{\mathbf{B}}_{13}, \widetilde{\mathbf{B}}_{21}, \widetilde{\mathbf{B}}_{22}, \widetilde{\mathbf{B}}_{23}, \widetilde{\mathbf{B}}_{31}, \widetilde{\mathbf{B}}_{32}, \widetilde{\mathbf{B}}_{33} \right]^T ,\\
		\mathbf{E} &\equiv \left[\mathbf{E}_{11}, \mathbf{E}_{12}, \mathbf{E}_{13}, \mathbf{E}_{21}, \mathbf{E}_{22}, \mathbf{E}_{23}, \mathbf{E}_{31}, \mathbf{E}_{32}, \mathbf{E}_{33} \right]^T ,\\
		\widetilde{\mathbf{E}} &\equiv \left[\widetilde{\mathbf{E}}_{11}, \widetilde{\mathbf{E}}_{12}, \widetilde{\mathbf{E}}_{13}, \widetilde{\mathbf{E}}_{21}, \widetilde{\mathbf{E}}_{22}, \widetilde{\mathbf{E}}_{23}, \widetilde{\mathbf{E}}_{31}, \widetilde{\mathbf{E}}_{32}, \widetilde{\mathbf{E}}_{33} \right]^T ,\\
		\boldsymbol{\Gamma} &\equiv \left[\boldsymbol{\Gamma}_1, \boldsymbol{\Gamma}_2, \boldsymbol{\Gamma}_3\right]^T ,\\
		\boldsymbol{\Delta} &\equiv \left[\boldsymbol{\Delta}_{11}, \boldsymbol{\Delta}_{22}, \boldsymbol{\Delta}_{33}, \boldsymbol{\Delta}_{23}, \boldsymbol{\Delta}_{13}, \boldsymbol{\Delta}_{12}\right]^T, \\
		\mathbf{F} &\equiv \boldsymbol{\Phi} + \boldsymbol{\Gamma} \mathbf{E}, \text{ and}\\
		\mathbf{G} &\equiv \boldsymbol{\Psi} + \boldsymbol{\Delta} \widetilde{\mathbf{E}},
		\end{align}
	\end{subequations}
	where $\mathbf{B}, \widetilde{\mathbf{B}} \in \mathbb{R}^{9N \times 9N}$, $\mathbf{E} \in \mathbb{R}^{9N \times 6N}$, $\widetilde{\mathbf{E}} \in \mathbb{R}^{9N \times 3N}$, $\boldsymbol{\Gamma} \in \mathbb{R}^{3N \times 9N}$, $\mathbf{F} \in \mathbb{R}^{3N \times 6N}$, $\mathbf{G} \in \mathbb{R}^{6N \times 3N}$ and $\boldsymbol{\Delta} \in \mathbb{R}^{6N \times 9N}$. In the presence of the C-PML, the single matrix equation to determine the updated wavefield variables after time $\Delta t$ is given by
	\begin{equation}
	\V^\prime_{m+1}=\mathbf{W'}\V^\prime_m,
	\end{equation} 
	where 
$
    \mathbf{v}_{m}^\prime = \begin{bmatrix}
	\overline{\boldsymbol{\phi}}_{m - \frac{1}{2}} \\
	\overline{\dot{\mathbf{u}}}_{m-\frac{1}{2}} \\
	\overline{\boldsymbol{\psi}}_{m} \\
	\overline{\boldsymbol{\sigma}}_m
	\end{bmatrix},
$
	and the propagator matrix $\mathbf{W}^\prime \in \mathbb{R}^{27N \times 27N}$ is given by 
	\begin{align}
	\mathbf{W}^\prime = \begin{bmatrix}
	\mathbf{B} & \mathbf{0}_{9N \times 3N} & \mathbf{0}_{9N \times 9N} & \mathbf{E} \\
	\boldsymbol{\Gamma} \mathbf{B} & \mathbf{J} & \mathbf{0}_{3N \times 9N} & \mathbf{F} \\
	\widetilde{\mathbf{E}} \boldsymbol{\Gamma} \mathbf{B} & \widetilde{\mathbf{E}} \mathbf{J} & \widetilde{\mathbf{B}} & \widetilde{\mathbf{E}} \mathbf{F} \\
	\mathbf{G} \boldsymbol{\Gamma} \mathbf{B} & \mathbf{G} \mathbf{J} & \boldsymbol{\Delta} \widetilde{\mathbf{B}} & \mathbf{I}_{6N \times 6N} + \mathbf{G} \mathbf{F}
	\end{bmatrix}.
	\end{align}
	
	Hence, the wavefield quantities can be propagated forward in time from $t = 0$ to $t = (M-1)\Delta t$ as 
	\begin{align}\label{eq:PropagatePML}
	\begin{bmatrix}
	&\mathbf{v'}_0&\\
	&\mathbf{v'}_1&\\
	&\vdots&\\
	&\mathbf{v'}_{M-1}&\\
	\end{bmatrix}
	= \mathbf{T'}_{M-1}\cdots\mathbf{T'}_{1}
	\begin{bmatrix}
	&\mathbf{v'}_0&\\
	&\mathbf{0}_{27N \times 1}&\\
	&\vdots&\\
	&\mathbf{0}_{27N \times 1}&\\
	\end{bmatrix},
	\end{align}
	where the $27NM \times 27NM$ matrices $\mathbf{T'}_m (m= 1,...,M-1)$ are defined in terms of $\mathbf{W'}$ as 
	\begin{align}
	\mathbf{T'}_m \equiv 
	\begin{bmatrix}
	\mathbf{I}_{27N \times 27N}&\cdots&\mathbf{0}_{27N \times 27N}&\ \ \\
	\vdots&\ddots&\vdots&\mathbf{0}_{(m+1)\cdot 27N \times (M-m)\cdot 27N}\\
	\mathbf{0}_{27N \times 27N}&\cdots&\mathbf{I}_{27N \times 27N}&\ \ \\
	\mathbf{0}_{27N \times 27N}&\cdots&\mathbf{W'}&\ \ \\
	\mathbf{0}_{(M-m-1)\cdot 27N \times m\cdot 27N}&\ &\ &\mathbf{0}_{(M-m-1)\cdot 27N \times (M-m)\cdot27N}
	\end{bmatrix},
	\end{align}
	with $\mathbf{W'}$ residing between the $(27N(m-1) + 1)^{th}$ to $27Nm^{th}$ column and the $(27Nm + 1)^{th}$ to $27N(m+1)^{th}$ rows of $\mathbf{T'}_m$.
	From the equation of state in Eqn.~\cref{eq:subeq1},~\cref{eq:subeq2} and the initial conditions in Eqn.~\cref{eq:Ini}, the vector $(\mathbf{v}'_0,\mathbf{0}_{27N \times 1},\cdots,\mathbf{0}_{27N \times 1})^T$ can be computed from the initial pressure distribution $\mathbf{p}_0$ as 
	\begin{align}\label{eq:initialp0PML}
	\begin{bmatrix}
	&\mathbf{v}'_0&\\
	&\mathbf{0}_{27N \times 1}&\\
	&\vdots&\\
	&\mathbf{0}_{27N \times 1}&\\
	\end{bmatrix} =
	\mathbf{T}'_0\mathbf{p}_0
	,
	\end{align}
	\text{ where } 
	\begin{align}
	\mathbf{T}'_0 &\equiv 
	\begin{bmatrix}
	\boldsymbol{\tau}',
	 \mathbf{0}_{27N\times N} \cdots,\mathbf{0}_{27N\times N}
	\end{bmatrix}^T \in \mathbb{R}^{27NM \times N} \text{ and }\\
	\boldsymbol{\tau}' &\equiv
	\begin{bmatrix}
		\mathbf{0}_{21N\times N},-\mathbf{I}_{N \times N},-\mathbf{I}_{N \times N},-\mathbf{I}_{N \times N},\mathbf{0}_{3N\times N}
	\end{bmatrix}^T \in \mathbb{R}^{27N \times N}
	\end{align}
	and $\mathbf{p}_0$ is defined by Eqn.~\cref{eq:p0}.  Note that the memory variable associated with the spatial derivative of the stress tensor is set to zero initially. This implies that the support of the initial pressure distribution should be at least 2 grid positions away from the PML in all three directions. 
	
	In general, the transducer locations $\mathbf{r}_l^d$ at which the data $\mathbf{\hat{p}}$ are recorded will not coincide with the vertices of the Cartesian grid at which the propagated field quantities are computed. The measured data $\mathbf{\hat{p}}$ can be related to the computed field quantities via an interpolation operation defined as 
	\begin{align}\label{eq:interpgridPML}
	\mathbf{\hat{p}} = \mathbf{M'} \begin{bmatrix}
	&\mathbf{v}'_0&\\
	&\mathbf{v}'_{1}&\\
	&\vdots&\\
	&\mathbf{v}'_{M -1}&\\
	\end{bmatrix},
	\text{ where } 
	\mathbf{M'} \equiv \begin{bmatrix}
	\boldsymbol{\Theta}'&\mathbf{0}_{L \times 27N}&\cdots&\mathbf{0}_{L \times 27N}\\
	\mathbf{0}_{L \times 27N}&\boldsymbol{\Theta}'&\cdots&\mathbf{0}_{L \times 27N}\\
	\vdots&\vdots&\ddots&\vdots\\
	\mathbf{0}_{L \times 27N}&\mathbf{0}_{L \times 27N}&\cdots&\boldsymbol{\Theta}'
	\end{bmatrix} \in \mathbb{R}^{LM \times 27N}.
	\end{align}
	Here, $\boldsymbol{\Theta}' \equiv \begin{bmatrix}
	s'_1,\cdots,s'_L
	\end{bmatrix}^T \in \mathbb{R}^{L \times 27N}$, where $l= 1,\cdots,L$ and \begin{align}s'_l = \begin{bmatrix} \mathbf{0}_{1\times 21N}, -\mathbf{R}_l,-\mathbf{R}_l, -\mathbf{R}_l,\mathbf{0}_{1\times 3N}  \end{bmatrix}\end{align} is a $1  \times 27N$ row vector. The elements of row vector $\mathbf{R}_l \in \mathbb{R}^{1\times N}$ are assigned values to compute the pressure at the $l^{th}$ transducer using trilinear interpolation.
	Thus, from Eqn.~\cref{eq:PropagatePML},~\cref{eq:initialp0PML},~\cref{eq:interpgridPML}, the explicit form of the system matrix that solves the initial value problem of Eqn.~\cref{eq:Elastic} in a discrete setting with a C-PML is given by
	\begin{align}\label{eq:STforHPML}
	\mathbf{\hat{p}} = \mathbf{M'}\mathbf{T'}_{M-1}\cdots\mathbf{T'}_1\mathbf{T'}_0\mathbf{p}_0.
	\end{align}
	Comparing Eqn.~\cref{eq:STforHPML} with  Eqn.~\cref{eq:Imageeqn} the explicit form of the system matrix is given by 
	\begin{align}\label{eq:HPML}
	\mathbb{H} = \mathbf{M'}\mathbf{T'}_{M-1}\cdots\mathbf{T'}_1\mathbf{T'}_0.
	\end{align}
	Hence, the explicit form of $\mathbb{H}^{\dagger}$ is given by
	\begin{align}\label{eq:adjHPML}
	\mathbb{H}^{\dagger} = \mathbf{T'}_0^{\dagger}\mathbf{T'}_1^{\dagger}\cdots \mathbf{T'}_{M-1}^{\dagger}\mathbf{M'}^{\dagger}.
	\end{align}
	The action of the adjoint matrix on the measured pressure data $\mathbf{\hat{p}}$ was implemented according to Eqn.~\cref{eq:adjHPML}. The state equations for computing $\mathbf{p}^{adj} = \mathbb{H}^{\dagger}\mathbf{\hat{p}}$ with the incorporation of the C-PML can be written as,
	\begin{subequations}
	\begin{align}
	\mathbf{v'}_{M-1} &= \boldsymbol{\Theta}'^{T}\mathbf{\hat{p}}_{M-1}\\
	\mathbf{v'}_{m} &= \boldsymbol{\Theta}'^{T}\mathbf{\hat{p}}_{m - 1} + \mathbf{W'}^{T}\mathbf{v}'_{m} \nonumber \\
	&m = M-1,\cdots,1 \label{eq:AdjupPML}\\
	\mathbf{p}^{adj} &= \boldsymbol{\tau}'^{T}\mathbf{v}'_0.
	\end{align} 
	\end{subequations}
	Similar to Eqn.~\cref{eq:backwardstep}, the recursive temporal backward update step of Eqn.~\cref{eq:AdjupPML} can be written as 
		\begin{subequations}
			\begin{align}
			\tilde{\boldsymbol{\psi}}_{m-1} &= \overline{\boldsymbol{\psi}}_{m+1} + \boldsymbol{\Delta}^\dagger \overline{\boldsymbol{\sigma}}_{m+1} \\
			\tilde{\dot{\mathbf{u}}}_{m-\frac{1}{2}} &= \overline{\dot{\mathbf{u}}}_{m+\frac{1}{2}} + \tilde{\mathbf{E}}^\dagger \overline{{\Psi}}_{m+1} + \mathbf{G}^\dagger \overline{\boldsymbol{\sigma}}_{m+1} \\
			\tilde{\boldsymbol{\phi}}_{m-\frac{1}{2}} &= \overline{\boldsymbol{\phi}}_{m + \frac{1}{2}} + \boldsymbol{\Gamma}^\dagger \tilde{\dot{\mathbf{u}}}_{m-\frac{1}{2}} \\
			\overline{\boldsymbol{\sigma}}_{m} &= \overline{\boldsymbol{\sigma}}_{m+1} + \mathbf{E}^\dagger \overline{\boldsymbol{\phi}}_{m + \frac{1}{2}} + \mathbf{F}^\dagger \tilde{\dot{\mathbf{u}}}_{m-\frac{1}{2}} + \boldsymbol{\mathcal{I}}^\prime_4 \boldsymbol{\Theta}'^T \hat{\mathbf{p}}_{m+1}  \\
			\overline{\dot{\mathbf{u}}}_{m-\frac{1}{2}} &= \mathbf{J} \tilde{\dot{\mathbf{u}}}_{m-\frac{1}{2}} \\
			\overline{\boldsymbol{\psi}}_{m-1} &= \tilde{\mathbf{B}} \tilde{\boldsymbol{\psi}}_{m-1} \\
			\overline{\boldsymbol{\phi}}_{m-
				\frac{1}{2}} &= \mathbf{B} \tilde{\boldsymbol{\phi}}_{m-\frac{1}{2}},
			\end{align}
		\end{subequations}
		where $\boldsymbol{\mathcal{I}}^\prime_4 \mathbf{v}^\prime_m \equiv \overline{\boldsymbol{\sigma}}_m$.

\section*{Acknowledgments}
This work was supported in part by NIH award EB01696301, \\ 5T32EB01485505
and NSF award DMS1614305.

\bibliographystyle{siamplain}
\bibliography{references}
\end{document}